\RequirePackage[dvipsnames,prologue]{xcolor}
\documentclass[dvipsnames, prologue, sigconf, screen]{acmart}
\definecolor{green}{RGB}{50,205,50}
\definecolor{blue}{RGB}{135,206,235}
\definecolor{purple}{RGB}{204,204,255}
\definecolor{indigo}{RGB}{216,191,216}
\definecolor{red}{RGB}{250,128,114}
\definecolor{yellow}{RGB}{218,165,32}
\usepackage{csquotes}
\usepackage{listings}
\usepackage{tabularx}
\usepackage{longtable}
\MakeOuterQuote{"}
\AtBeginDocument{%
  \providecommand\BibTeX{{%
    \normalfont B\kern-0.5em{\scshape i\kern-0.25em b}\kern-0.8em\TeX}}}

\copyrightyear{2024}
\acmYear{2024}
\setcopyright{rightsretained}
\acmConference[CHI '24]{Proceedings of the CHI Conference on Human Factors in Computing Systems}{May 11--16, 2024}{Honolulu, HI, USA}
\acmBooktitle{Proceedings of the CHI Conference on Human Factors in Computing Systems (CHI '24), May 11--16, 2024, Honolulu, HI, USA}
\acmDOI{10.1145/3613904.3641920}
\acmISBN{979-8-4007-0330-0/24/05}

\usepackage[dvipsnames]{xcolor}
\colorlet{RED}{red}
\newcommand{\rev}[1] {{#1}}

\begin{document}

\title{Jigsaw: Supporting Designers to Prototype Multimodal Applications by Chaining AI Foundation Models}

\author{David Chuan-En Lin}
\affiliation{%
  \institution{Carnegie Mellon University}
  \streetaddress{5000 Forbes Ave.}
  \city{Pittsburgh, PA}
  \country{USA}
  }
\email{chuanenl@cs.cmu.edu}

\author{Nikolas Martelaro}
\affiliation{%
  \institution{Carnegie Mellon University}
  \streetaddress{5000 Forbes Ave.}
  \city{Pittsburgh, PA}
  \country{USA}
  }
\email{nikmart@cmu.edu}

\renewcommand{\shortauthors}{David Chuan-En Lin and Nikolas Martelaro}

\begin{abstract}
  Recent advancements in AI foundation models have made it possible for them to be utilized off-the-shelf for creative tasks, including ideating design concepts or generating visual prototypes. However, integrating these models into the creative process can be challenging as they often exist as standalone applications tailored to specific tasks. To address this challenge, we introduce Jigsaw, a prototype system that employs puzzle pieces as metaphors to represent foundation models. Jigsaw allows designers to combine different foundation model capabilities across various modalities by assembling compatible puzzle pieces. To inform the design of Jigsaw, we interviewed ten designers and distilled design goals. In a user study, we showed that Jigsaw enhanced designers' understanding of available foundation model capabilities, provided guidance on combining capabilities across different modalities and tasks, and served as a canvas to support design exploration, prototyping, and documentation.
\end{abstract}

\ccsdesc[500]{Human-centered computing~Human computer interaction (HCI)}
\ccsdesc[500]{Applied computing~Arts and humanities}

\keywords{prototyping, machine learning, foundation models, multimodal, visual programming interface}


\begin{teaserfigure}
  \centering
  \includegraphics[width=16cm]{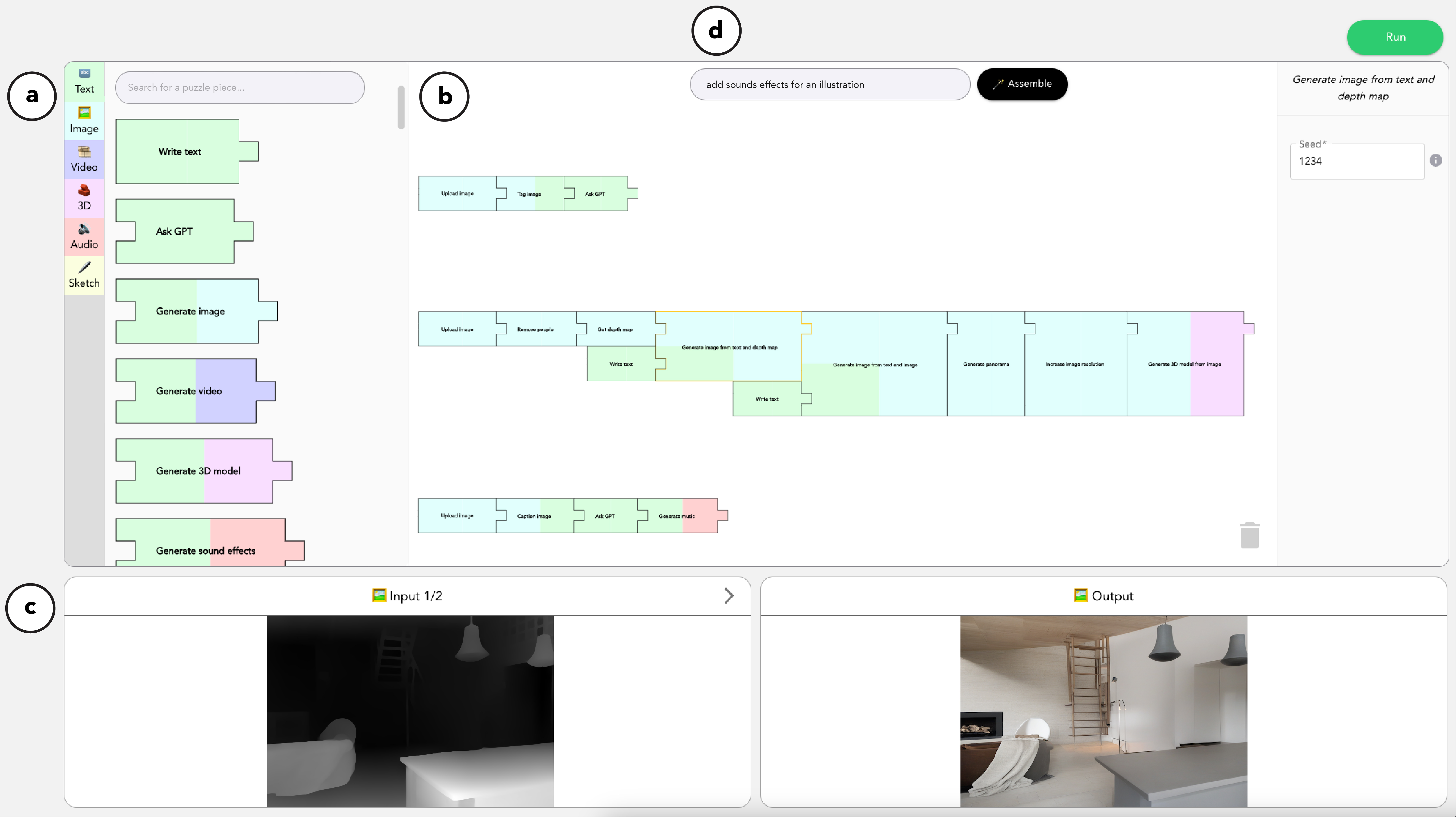}
  \caption{\rev{Jigsaw is a prototyping system that lets designers generate creative content with AI foundation models represented as puzzle pieces. Designers can first search for model capabilities via the Catalog Panel (a). Designers can drag corresponding puzzle pieces onto the Assembly Panel (b) and combine different AI capabilities across varying modalities into a chain by assembling compatible pieces. Designers can specify inputs and observe the intermediate results of a chain via the Input and Output Panels (c). Designers can ask the Assembly Assistant (d) to recommend a chain of AI models to accomplish a task.}}
  \Description{Jigsaw is a prototyping system that lets designers generate creative content with AI foundation models represented as puzzle pieces. Designers can first search for model capabilities via the Catalog Panel (a). Designers can drag corresponding puzzle pieces onto the Assembly Panel (b) and combine different AI capabilities across varying modalities into a chain by assembling compatible pieces. Designers can specify inputs and observe the intermediate results of a chain via the Input and Output Panels (c). Designers can ask the Assembly Assistant (d) to recommend a chain of AI models to accomplish a task.}
  \label{fig:teaser}
\end{teaserfigure}

\maketitle

\section{Introduction}

The past year has seen substantial progress in the capabilities of AI foundation models \cite{bommasani2021opportunities}. These models, which are pre-trained on vast quantities of data, can perform many tasks ``off the shelf'' without further training. Consequently, many foundation models essentially become input-output systems, simplifying the complexities of working with AI by abstracting models to their core capability \cite{yildirim2023creating}. Until recently, creating an AI-enabled system necessitated users to curate their own data, train a model, and occasionally modify the model architecture to adapt to their use cases \cite{transfer-learning}.

With powerful plug-and-play capabilities, many designers have begun embracing AI foundation models to enhance their creative workflows. New foundation models support a wide variety of tasks and modalities, including large language models \cite{wei2022emergent} such as GPT \cite{brown2020language} for text generation and processing, image generation models such as Stable Diffusion \cite{rombach2022high}, image segmentation models such as Segment Anything \cite{kirillov2023segment}, and models for video \cite{wang2023modelscope}, 3D model \cite{jun2023shap}, audio \cite{kreuk2022audiogen} generation. 

However, despite the variety of capabilities offered, the integration of foundation models within the creative process can be challenging. Our initial observations suggest that these models are often used for one-off tasks or as standalone applications. For instance, a designer might use ChatGPT \cite{chatgpt} to brainstorm and generate ideas, or they might use Midjourney \cite{midjourney} to generate visual prototypes. To incorporate the results of these models into their broader creative process, designers manually copy and paste the results into another design tool. Moreover, despite the variety of available models, designers typically only use a small selection of highly publicized models (ChatGPT, Stable Diffusion, MidJourney) and are often unaware of the range of capabilities and modalities they could potentially utilize from lesser-known models.

To gain a deeper understanding of designers' challenges when using current AI models in their creative processes, we conducted a formative study with ten designers. From our formative study, we identified four key challenges:
\begin{enumerate}
    \item Designers are often unaware of the full range of capabilities offered by different types of foundation models.
    \item Designers struggle with the need to be "AI-friendly," which includes difficulties in forming effective prompts and selecting optimal parameters.
    \item Designers find it challenging to cross-integrate foundation models that exist on different platforms and are specialized for different modalities.
    \item Designers find prototyping with these models to be a slow and arduous process.
\end{enumerate}

Based on the findings from the formative study, we derived four design goals, which informed the development of Jigsaw, a block-based prototype system that represents foundation models as puzzle pieces and allows designers to combine the capabilities of different foundation models by assembling compatible puzzle pieces together. 
Jigsaw includes features that help designers discover available foundation model capabilities and find the right model for their use case.
Jigsaw also includes "glue" puzzle pieces that translate design ideas into prompts for other models, clear explanations of parameters to help users make model adjustments, and an Assembly Assistant that recommends potential combinations of foundation models to accomplish a task specified by the designer. 
To assess the utility of Jigsaw, we invited ten designers from the formative study to test the system. We evaluate how well designers create creative AI workflows given a design brief and during free exploration. 
The results show that Jigsaw helps designers better understand the capabilities offered by current foundation models, provides intuitive mechanisms for using and combining models across diverse modalities, and serves as a visual canvas for design exploration, prototyping, and documentation.

This research thus contributes:
\begin{itemize}
\item{A \textbf{formative study} with ten designers that identifies the challenges designers face when using AI foundation models to support their work.}
\item{\textbf{Jigsaw}, a prototype system that assists designers in combining the capabilities of AI foundation models across different tasks and modalities through assembling compatible puzzle pieces.}
\item{A \textbf{user study} that demonstrates the utility of Jigsaw to designers and informs areas for future block-based prototyping systems for prototyping with AI foundation models.}
\end{itemize}

\section{Related Work}

This work draws on prior research in AI foundation models\rev{, }visual programming interfaces\rev{, and designer-AI interaction}.

\subsection{AI Foundation Models}

The term "foundation models" characterizes an emerging family of machine learning models \cite{bommasani2021opportunities}, often underpinned by the Transformer architecture \cite{vaswani2017attention} and trained on vast amounts of data. The researchers who introduced this term defined foundation models as "models trained on broad data (generally using self-supervision at scale) that can be adapted to a wide range of downstream tasks." \cite{foundation-models} The strength of foundation models lies in their capacity for out-of-the-box usage across various tasks. This signifies an improvement from the previous AI landscape, where users had to create their own datasets for custom use cases and fine-tune models \cite{transfer-learning}.

Prominent examples of foundation models include large language models such as GPT \cite{brown2020language} which can perform a variety of text generation tasks and image generation models such as Stable Diffusion \cite{rombach2022high} which can generate a diverse range of images from text-based prompts. 
Foundation models also go beyond generative models and include models for tasks such as classification \cite{radford2021learning}, detection \cite{zhou2022detecting}, segmentation \cite{kirillov2023segment}, spanning a range of modalities including text \cite{brown2020language}, image \cite{rombach2022high}, video \cite{wang2023modelscope}, 3D models \cite{jun2023shap}, and audio \cite{kreuk2022audiogen}. 
Many foundation models perform tasks \textit{across modalities}, such as text-to-\textit{x} generative models and \textit{x}-to-text classification models.
In turn, this allows foundation models to be treated as \textit{x}-to-\textit{x} input-output systems.
Such abstraction greatly simplifies how people can use and combine such models in larger AI-enabled systems.
Our research aims 1) to inform designers about the capabilities offered by foundation models that can be useful for creative tasks, and 2) to incorporate these capabilities into their creative workflows. In particular, we are interested in exploring how designers can \textit{combine} the capabilities of multiple models \textit{across different tasks and modalities} by connecting them together on a visual interface.

\subsection{Visual Programming Interfaces}

Visual programming interfaces (VPIs) have been extensively studied as tools to aid users in designing and implementing systems through graphical elements rather than text-based code \cite{myers1990taxonomies}. A key benefit of VPIs is their lower entry barrier for novice programmers \cite{whitley1997visual}. There are primarily two main paradigms for VPIs. The first, the dataflow paradigm, lets users specify how a program transforms data from step to step by connecting nodes in a directed graph. Pioneering work in this area includes Prograph \cite{cox1989prograph} and LabVIEW \cite{kodosky2020labview}. 
The second paradigm utilizes block-based function representations and lets users create programs by connecting compatible components together. Notable works in this area include Scratch \cite{resnick2009scratch} and Blockly \cite{fraser2013blockly}. Many commercial creative applications have adopted VPIs, including game engines such as Unity \cite{unity}, CAD tools such as Grasshopper \cite{grasshopper}, and multimedia development tools such as Max/MSP \cite{max}.

VPI concepts have been applied to machine learning applications. For example, \textit{Teachable Machine} \cite{carney2020teachable} uses a visual interface to help students learn to train a machine learning model. 
\textit{ML Blocks} \cite{williams2022ml} assists developers in training, evaluating, and exporting machine learning model architectures.
\rev{Very recently, researchers in both academia and industry have worked on VPIs that support the creation of AI workflows through the combination of pre-trained models.
Several works have investigated node-based interfaces for building Large Language Model (LLM) pipelines, including \textit{PromptChainer} \cite{wu2022promptchainer}, \textit{FlowiseAI} \cite{flowiseai}, and \textit{Langflow} \cite{langflow}.
Most closely related to our work are \textit{Rapsai} by Du et al. \cite{du2023rapsai} and \textit{ComfyUI} \cite{comfyui}.
Both tools provide a node-based interface for machine learning researchers and enthusiasts to build multimedia machine learning pipelines.
These tools are catered more toward users with at least some background knowledge in AI programming, giving users the flexibility to customize the tools through programming at the expense of exposing more technical elements to the user.}

\rev{Our work builds upon prior and concurrent VPI tools and research.
However, we made several design choices for our tool to help better support non-technical designers (Table \ref{tab:related-tools}).
First, our tool leverages a block-based VPI paradigm, which has been shown to be effective in supporting novice programming learners \cite{resnick2009scratch}.
Second, in the same spirit as other creative AI tools such as \textit{RunwayML} \cite{runwayml}, our tool supports AI capabilities of a diverse range of modalities.
Third, our tool offers integrated AI assistance features for designers, such as the Assembly Assistant (Section \ref{section:assembly-assistant}), semantic search (Section \ref{section:semantic-search}), and glue pieces (Section \ref{section:glue}). We build on recent advances in the reasoning capabilities of LLMs to power these features \cite{wu2023visual}.
}
To the best of our knowledge, this research is the first to study 1) supporting \textit{non-technical designers} in prototyping design workflows with AI through a block-based visual interface and 2) utilizing the \textit{plug-and-play capabilities of AI foundation models} that have emerged over the past year, covering a \textit{diverse range of tasks and modalities}.

\begin{table*}[]
\caption{\rev{Comparison of Jigsaw against related tools. Jigsaw supports non-technical designers with a beginner-friendly block editor and offers AI capabilities across multiple modalities. Jigsaw's Assembly Assistant can help automatically recommend a chain of AI models for a designer-specified task.}}
\label{tab:related-tools}
\resizebox{\textwidth}{!}{%
\begin{tabular}{|l|l|l|l|l|l|}
\hline
\textbf{Tool name} &
  \textbf{Interface design} &
  \textbf{Supported modalities} &
  \textbf{Target audience} &
  \textbf{Assembly assistant} &
  \textbf{Release year} \\ \hline
Jigsaw &
  \textbf{Block editor} &
  \textbf{Text, Image, Video, 3D, Audio, Sketch} &
  \textbf{Designers} &
  \textbf{Yes} &
  2023 \\ \hline
PromptChainer  & Node editor       & Text                          & Developers         & No           & 2022 \\ \hline
FlowiseAI       & Node editor       & Text                          & Developers         & No           & 2023 \\ 
\hline
Langflow       & Node editor       & Text                          & Developers         & No           & 2023 \\ \hline
Rapsai         & Node editor       & Text, Image                   & ML researchers     & No           & 2023 \\ \hline
ComfyUI        & Node editor       & Text, Image, Video            & ML enthusiasts         & No           & 2023 \\ \hline

RunwayML       & Standalone models & Text, Image, Video, 3D, Audio & \textbf{Designers} & No           & 2018 \\ \hline
Visual ChatGPT & Chatbot           & Text, Image                   & General public         & \textbf{Yes} & 2023 \\ \hline 
\end{tabular}%
}
\end{table*}

\subsection{\rev{Designer-AI Interaction}}

\rev{Several works from the HCI design community have examined the ways in which designers perceive and interact with AI. Chiou et al. \cite{chiou2023designing} follow a Research through Design (RtD) \cite{zimmerman2007research} approach and find that AI can offer designers new perspectives and avenues of design exploration. Shi et al. \cite{shi2023understanding} conduct a landscape analysis of AI and suggest the opportunity to build more tools that enable co-creativity between designers and AI. Yang \cite{yang2018machine} proposes the vision of designers working with AI as a "design material". This research follows this thread of work to build a tool to help designers prototype new design workflows using AI and with the support of AI.}

\rev{Subramonyam et al. \cite{subramonyam2021towards} argue that a challenge with using AI as a design material is that the properties of AI only emerge as part of user experience design. They thus employ data probes with user data to help elicit AI properties and facilitate working with AI as a design material. Yang et al. \cite{yang2020re} identify that designers often find designing with AI difficult due to uncertainty about the AI’s capabilities and the complexity of the AI’s outputs. Gmeiner et al. \cite{gmeiner2023exploring} identify the primary challenges for designers when co-creating with AI design tools as understanding and manipulating AI outputs and communicating design goals to the AI. In this research, we offer mechanisms to help designers overcome these challenges, such as conveying AI capabilities, supporting easy inspection and manipulation of AI outputs with real data, and allowing users to communicate design goals to the AI using natural language.}

\rev{Liu et al. \cite{liu2022design} find that when designers use the "right" prompts, they achieve significantly higher quality results from generative models. However, Zamfirescu et al. \cite{zamfirescu2023johnny} find that people generally struggle with writing effective prompts. In this research, we introduce a puzzle piece (translation glue) to help designers automatically translate pieces of text into prompts. Yang et al. \cite{yang2018investigating} find that designers are more successful when they collaborate with data scientists. Using RtD, Yildirim et al. \cite{yildirim2022experienced} identify that designers develop boundary objects to communicate design intentions with data scientists. In this research, we let designers document their creative process on a canvas (Assembly panel), which designer participants in our user study found to be a useful boundary object for sharing and explaining ideas.}

\section{Formative Study}
We conducted a formative interview study with ten designers to understand how designers attempt to use AI in their work and inform the development of a new tool to support creative work with AI.

\subsection{Participants and Procedure}
We interviewed ten designers (P1-P10, 6 male and 4 female, aged 24-39), recruited through known contacts and word of mouth. The designers come from diverse specializations, such as interior design, product design, graphic design, and video game design. All participants have more than five years of design work experience and use AI tools to support aspects of their design processes, such as ChatGPT, Midjourney, and DALL-E \cite{ramesh2021zero}.
We conducted one-hour interviews remotely over video conferencing, asking participants to describe their typical creative workflow, how they use AI to support their work, the specific AI tools they use, and the pain points they face using AI.
Following the interviews, the first author conducted a thematic analysis of interviews and summarized participants' key challenges.

\subsection{Findings and Discussion}
We identify four key challenges designers face when using AI to support their work.

\subsubsection{\rev{C1: Limited Knowledge of AI Capabilities}}
Despite the broad spectrum of AI foundation models available, designers felt they had limited knowledge of existing models and their capabilities. As a result, they felt like they were underutilizing the creative support these models could provide. Designers found it challenging to \textit{"understand the capabilities of various models in a crowded market (P1)"}, making it difficult to determine which model is most suitable for a specific task. Additionally, designers expressed a desire to \textit{"easily view a few example results from the models (P5)"}, which would allow them to quickly assess the model's capabilities and determine if its results align with their intended use case.

\subsubsection{\rev{C2: Tedious to be AI-friendly}}
After deciding on an AI model to use, designers stated that it is challenging to be "AI-friendly." This includes crafting effective prompts (for generative models) and setting optimal parameter values of AI models to ensure good results. Designers stated it can be time-consuming to \textit{"master the art of prompt creation (P2)"}, often dedicating a significant amount of time to simply translating their design idea into a functional prompt. As P5 stated, \textit{"behind every stunning image generated by Stable Diffusion lies a designer's patience and a relentless pursuit of the right prompt."} Moreover, our participants were often confused about different model parameters and how they affect the model's results, leading to \textit{"endless parameter tweaks (P10)"}.

\subsubsection{\rev{C3: Difficult to Combine Multiple Models}}
Designers felt that current models predominantly cater to simple and singular functionalities. Designers commented that for realistic design workflows, which involve multiple tasks and a range of modalities, they often find themselves having to switch between distinct AI platforms. This fragmented the design process, and as P9 stated, \textit{"switching between AI platforms felt like needing a different kitchen gadget for every step in a recipe."} In addition, designers often face compatibility issues between models when attempting to combine them, leading to time-consuming troubleshooting. They commented that it would be beneficial to \textit{"clearly know which models are compatible with one another (P6)."}

\subsubsection{{C4: Slow Prototyping and Iteration}}
Designers noted that \textit{"seamless prototyping and iteration is crucial to the design process (P6)"}. However, when working with AI, designers frequently found it challenging to quickly build prototypes and view results. Setting up and switching models can be a lengthy process that inhibits rapid experimentation. Furthermore, when creating workflows that involve chaining models, designers often can only view the final result. This makes it difficult to understand how individual models affect the final result and can make it challenging to explain design decisions to clients without tangible intermediate outputs.

\subsection{Design Goals}
To tackle the challenges designers encounter when using AI, we distill four design goals.

\subsubsection{D1: Catalog of AI Foundation Models}
To help designers gain a better understanding of available AI foundation models, we aim to compile a catalog of existing models. For each model, we should provide straightforward explanations of its capabilities along with examples of their results. Furthermore, we should provide mechanisms for designers to easily find models that can accomplish the specific tasks they have in mind.

\subsubsection{D2: User-friendly instead of AI-friendly}
We should provide mechanisms that reduce the need for designers to adapt to the nuances of AI models. First, we should incorporate assisted prompting techniques to help designers translate design ideas into prompts. Second, we should explain model parameters in laypeople's terms, including how altering different values will impact model results.

\subsubsection{D3: Intuitive Interface for Combining Models}
We should provide designers with an interface that allows them to easily combine multiple task-specific foundation models across a wide range of modalities. The interface should visually present clear affordances of which models can be combined. In addition, we should provide an assistive tool for suggesting model combinations.

\subsubsection{D4: Facilitate Effective Prototyping}
Designers place significant importance on experimentation and iteration. We should make it effortless for designers to experiment with different model combinations and be able to easily view results. Furthermore, we should let designers view intermediate results within a chain of models to help diagnose errors and aid in communicating design ideas with clients.

\section{Jigsaw}
\label{section:implementation}

The following outlines Jigsaw's four major components: the (1) Catalog Panel, (2) Assembly Panel, (3) Input and Output Panels, and (4) Assembly Assistant. We then describe how a designer can use Jigsaw with an example interior design workflow.

\subsection{Catalog Panel}
The \textit{Catalog Panel} assists designers in selecting suitable models for their tasks with a catalog of foundation model components (D1).

\begin{figure}[tbp]
  \centering
  \includegraphics[width=6cm]{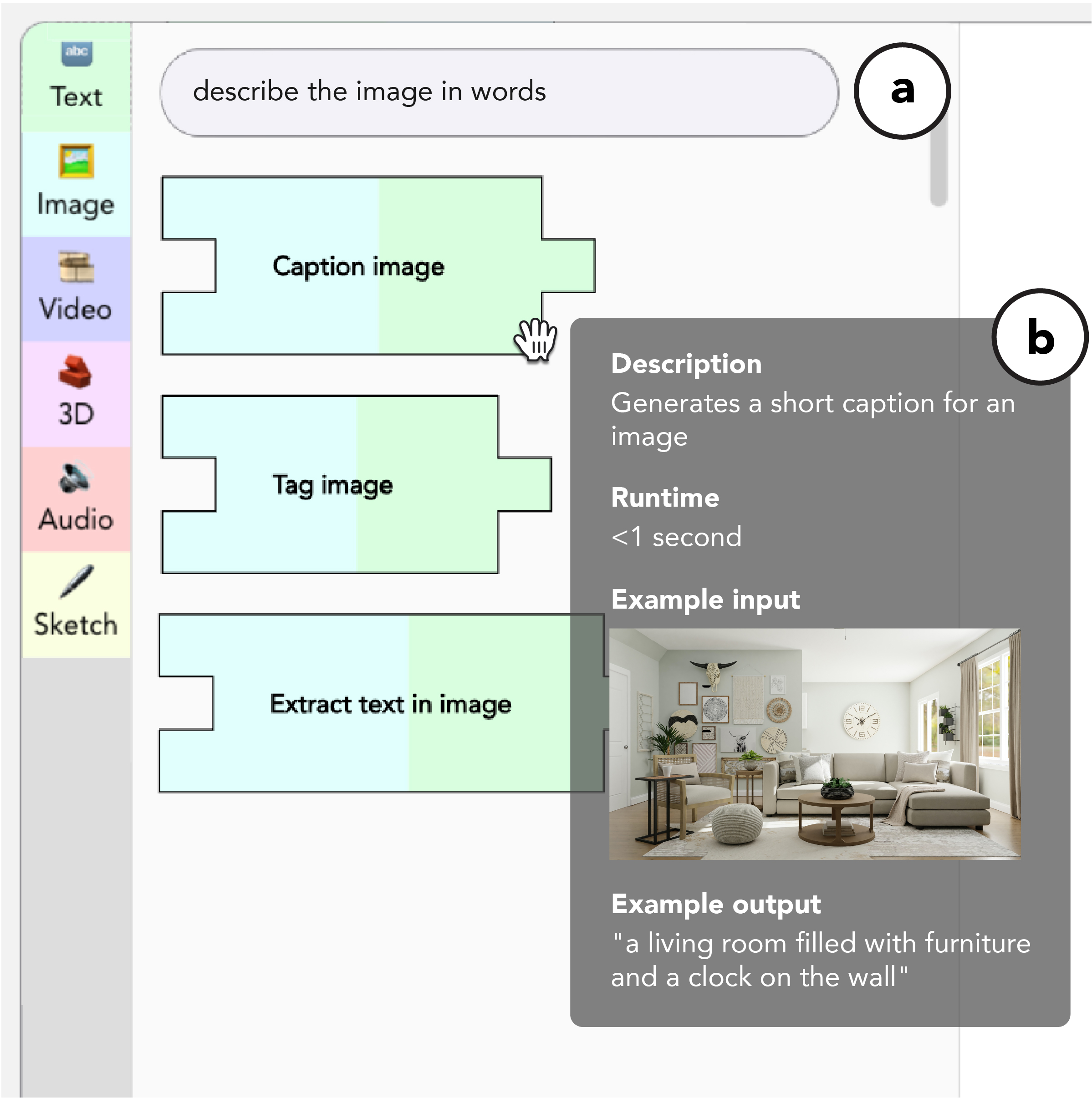}
  \caption{Users can search for model pieces by describing their task in the semantic search bar (a). Users can hover over a model piece to view a description of its capability, typical runtime, and an example input and output (b).}
  \Description{Users can search for model pieces by describing their task in the semantic search bar (a). Users can hover over a model piece to view a description of its capability, typical runtime, and an example input and output (b).}
  \label{fig:catalog-panel}
\end{figure}

\subsubsection{Curating a catalog of foundation models}
We identified six common modalities used in creative work, namely, text, image, video, 3D, audio, and sketches. Jigsaw curates available models across all possible pairwise permutations of modalities (e.g., \texttt{text-to-text}, \texttt{text-to-image}, \texttt{text-to-video}, ...). 
Jigsaw also includes foundation models that with dual input channels, such as ControlNet \cite{zhang2023adding}.
For tasks supported by multiple models, we prioritize models based on: 1) inference speed (ideally less than a minute to run), 2) zero-shot capability (plug-and-play use), and 3) the quality of results (models ranked highly on machine learning benchmarks). Overall, we implemented a catalog of thirty-nine models across six modalities (see Appendix \ref{appendix:models-list} for a full listing).

\subsubsection{Representing foundation models as puzzle pieces}
Considering foundational models as input/output systems, we represent them as puzzle pieces with input and output arms. There are two types of puzzle pieces: 1) the \textit{model} piece represents models with customizable parameters, and 2) the \textit{input} piece accepts input from the user via text, media file, or sketch (see Section \ref{section:input-output}). 
Jigsaw color-codes puzzle pieces based on their input and output modalities to signals what pieces are compatible with one another (D3). For example, a \texttt{text-to-image} piece would be colored green on the left and blue on the right.
When a user hovers over a puzzle piece, a tooltip provides a description of its capability, typical runtime, and an example of an input and output (D1).

\subsubsection{Helping users find the right piece}
\label{section:semantic-search}
Jigsaw provides two mechanisms for users to find model pieces for their tasks (D1): 1) puzzle pieces are grouped by input modality and 2) users can describe the task in the semantic search bar. The search returns model pieces with high semantic similarity to the query, scored using CLIP \cite{radford2021learning} text embeddings.

\begin{figure*}[tbp]
  \centering
  \includegraphics[width=9.2cm]{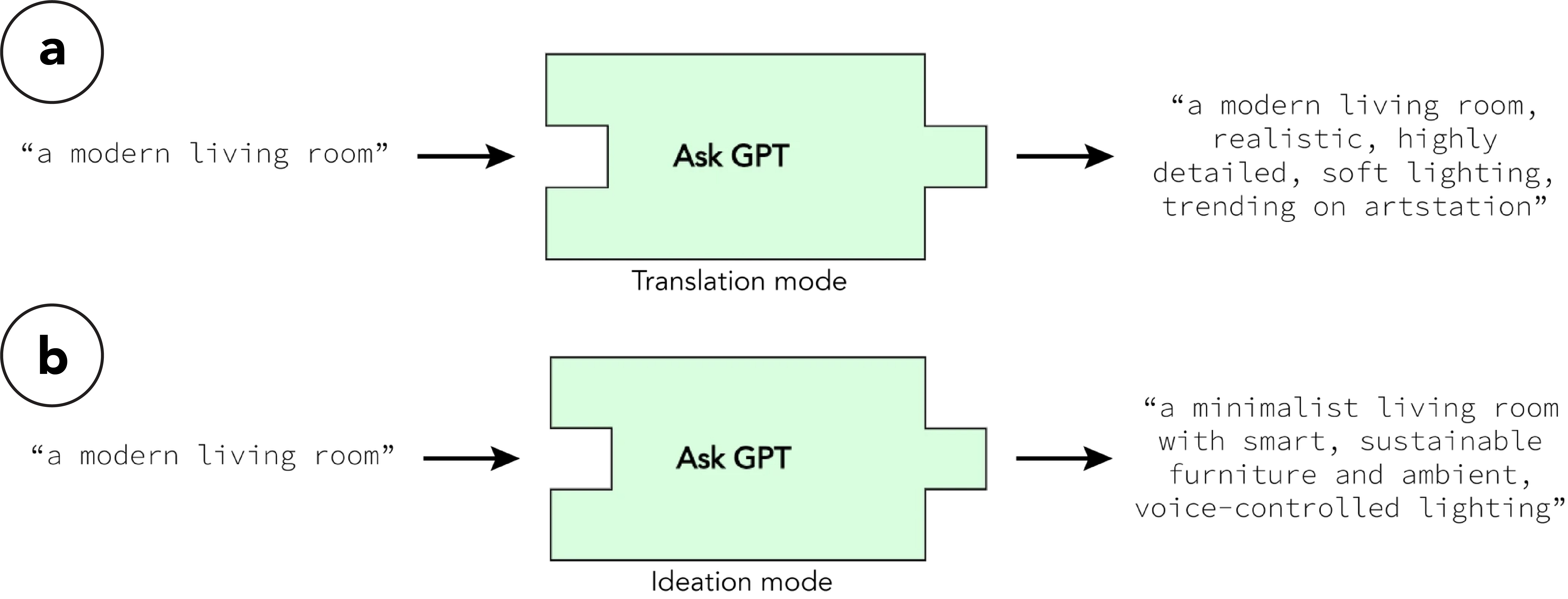}
  \caption{\rev{The \textit{translation} glue piece converts a piece of text into a prompt format suitable for text-to-\textit{x} generation models (a). The \textit{ideation} glue piece generates an idea for a design task (b).}}
  \Description{The translation glue piece converts a piece of text into a prompt format suitable for text-to-x generation models (a). The ideation glue piece generates an idea for a design task (b).}
  \label{fig:glue-pieces}
\end{figure*}

\subsubsection{LLMs as glue}
\label{section:glue}
There can be instances where models do not perfectly align, such as situations where intermediate reasoning is required. Drawing inspiration from \textit{Socratic Models} by Zeng et al. \cite{zeng2022socratic}, we utilize Large Language Models (LLMs) as connecting elements between model pieces. We refer to these instances as \textit{glue} pieces.
The user first attaches a model piece capable of conveying the content of a modality in text (\textit{x}-to-text).
Next, the user attaches the LLM glue piece for language-based reasoning (text-to-text).
Finally, the user attaches a model piece which translates text back into another modality (text-to-\textit{x}).
To help users connect model pieces in common use cases, Jigsaw includes three types of glue pieces:

\begin{enumerate}
    \item The \textit{custom} glue piece accepts any custom user instruction.
    \item The \textit{translation} glue piece converts a piece of text into a prompt that better aligns with text-to-\textit{x} models (e.g., Stable Diffusion) (D2) (Figure \ref{fig:glue-pieces}a). We ask GPT to transform an input into a prompt via the following prompt:

\begin{lstlisting}
Here are example prompts for a text-to-<modality> generation model: <list of example prompts>.
Transform <input data> into a prompt. Answer in only the transformed prompt.
\end{lstlisting}

\item The \textit{ideation} glue piece accepts a design task specified by the user and generates an idea (Figure \ref{fig:glue-pieces}b). We ask GPT to generate an idea via the following prompt.
\begin{lstlisting}
Generate an idea for <task> based on <input data>. Answer in one short sentence.
\end{lstlisting}
\end{enumerate}

\subsection{Assembly Panel}
\begin{figure*}[tbp]
  \centering
  \includegraphics[width=11cm]{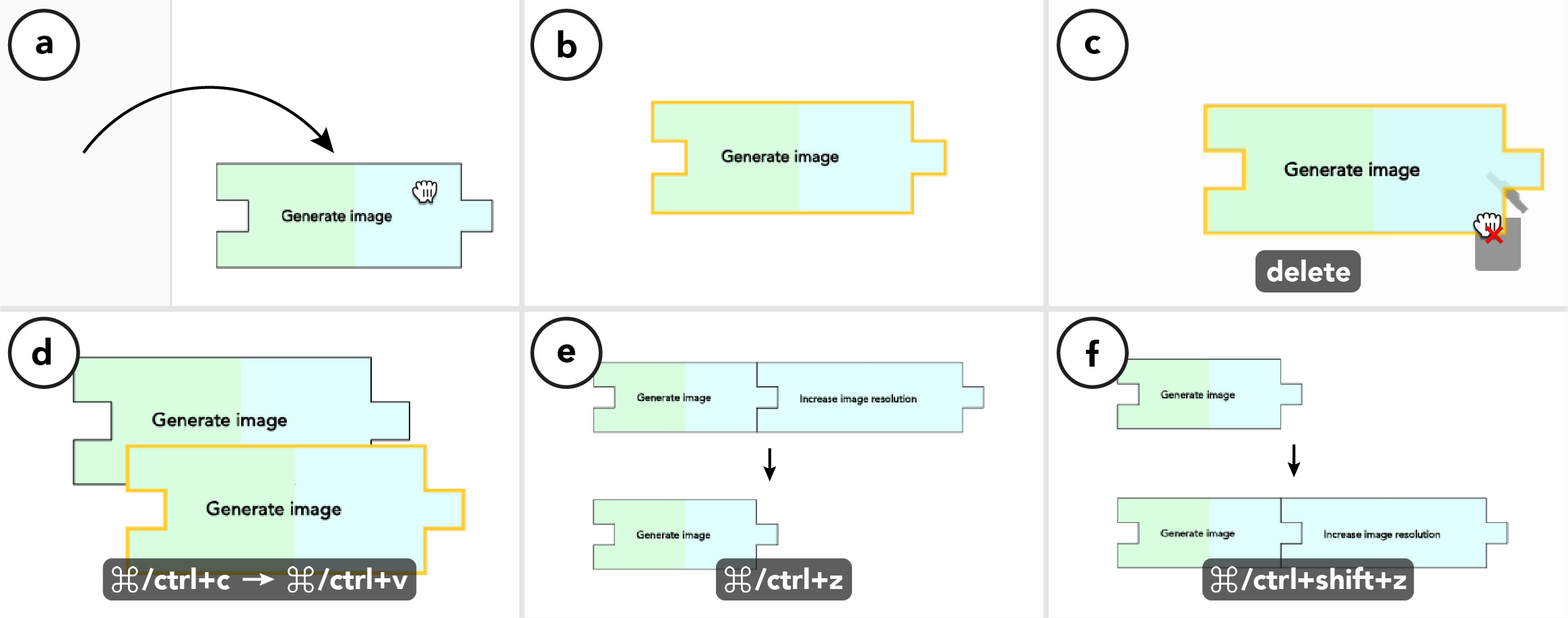}
  \caption{Users can drag puzzle pieces from the Catalog Panel onto the Assembly Panel (a), select pieces on the Assembly Panel by clicking on them (b), and remove pieces by dragging them to the trash bin or pressing the delete key (c). Users can duplicate pieces, and undo and redo actions using hotkeys (d-f).}
  \Description{Users can drag puzzle pieces from the Catalog Panel onto the Assembly Panel (a), select pieces on the Assembly Panel by clicking on them (b), and remove pieces by dragging them to the trash bin or pressing the delete key (c). Users can duplicate pieces, and undo and redo actions using hotkeys (d-f).}
  \label{fig:assembly-panel-operations}
\end{figure*}

The \textit{Assembly Panel} offers an infinite canvas for combining compatible foundation model puzzle pieces (D3) (Figure \ref{fig:assembly-panel-operations}).
When the user clicks on a model piece, a \textit{parameters sidebar} allows users to customize a model's specific parameters. 
Jigsaw pre-populates each model's parameters with default values that generally yield good results and defines limits so that the user can experiment with different values without the concern of breaking the model (D2).
Tooltips explain, in plain English, how the parameter influences model results and recommends optimal values for common scenarios.
Users can build multiple chains on the canvas and can run each chain separately, allowing parallel explorations and complex workflows.

\begin{figure*}[tbp]
  \centering
  \includegraphics[width=11cm]{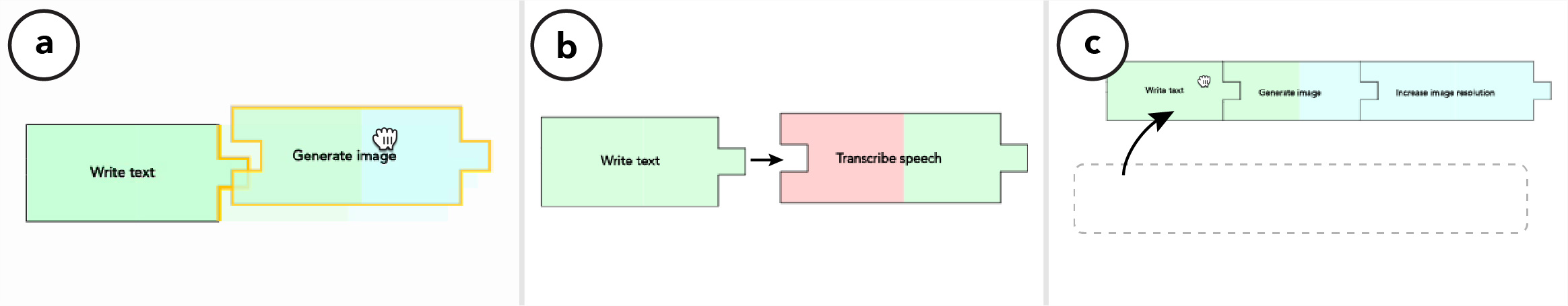}
  \caption{When the user drags a puzzle piece close to another compatible piece, Jigsaw displays a semi-transparent preview of the potential connection. If the user releases the puzzle piece, it will snap into place (a). Conversely, if the user attempts to connect a puzzle piece to an incompatible piece, the new piece will be repelled, ensuring that users do not force a fit (b). Users can move multiple puzzle pieces simultaneously (c).}
  \Description{When the user drags a puzzle piece close to another compatible piece, Jigsaw displays a semi-transparent preview of the potential connection. If the user releases the puzzle piece, it will snap into place (a). Conversely, if the user attempts to connect a puzzle piece to an incompatible piece, the new piece will be repelled, ensuring that users do not force a fit (b). Users can move multiple puzzle pieces simultaneously (c).}
  \label{fig:assembly-panel-interactions}
\end{figure*}

\subsection{Input and Output Panels}
\label{section:input-output}

\begin{figure*}[tbp]
  \centering
  \includegraphics[width=13cm]{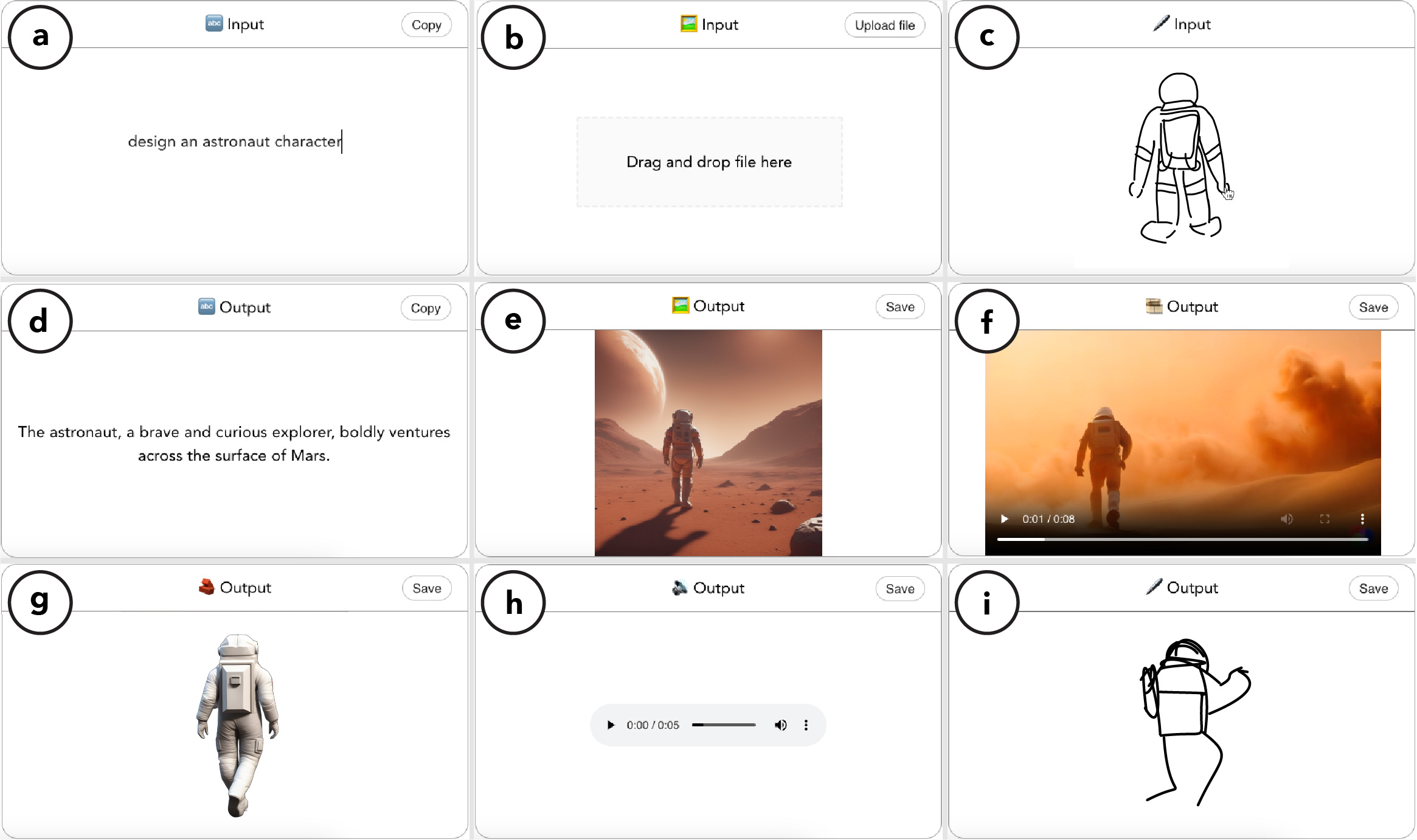}
  \caption{Text inputs can be directly typed into the Input Panel (a). Image, video, 3D, and audio inputs can be uploaded either by drag-and-drop or the file browser (b). Sketch inputs can be drawn (c). Text outputs can be viewed and copied by the user (d). Image, video, 3D, audio, and sketch outputs can be viewed in their respective media viewers and downloaded by the user (e-i).
}
  \Description{Text inputs can be directly typed into the Input Panel (a). Image, video, 3D, and audio inputs can be uploaded either by dragging and dropping or by accessing the file browser (b). Sketch inputs can be drawn (c). Text outputs can be viewed and copied by the user (d). Image, video, 3D, audio, and sketch outputs can be viewed in their respective media viewers and downloaded by the user (e-i)}
  \label{fig:input-output-panels}
\end{figure*}

The \textit{Input and Output Panels} allow users to input, view, and download media across modalities.
Users can type into the input panel (Figure \ref{fig:input-output-panels}a), upload files (Figure \ref{fig:input-output-panels}b), or draw sketches (Figure \ref{fig:input-output-panels}c). The Output Panel shows the result of the chain and lets users copy (Figure \ref{fig:input-output-panels}d) or download outputs (Figure \ref{fig:input-output-panels}e-i). 

The user can select a puzzle piece to view the intermediate inputs and outputs at that specific piece. This allows the user to observe how the data is transformed at each stage (D4). Additionally, within a chain, the user can view how the inputs for a puzzle piece affect the results of a puzzle piece located several steps downstream by holding the shift key to select multiple puzzle pieces.

\subsection{Assembly Assistant}
\label{section:assembly-assistant}
The \textit{Assembly Assistant} recommends a chain of puzzle pieces for a user-specified task. The designer would first provide a natural language description of a task, such as ``Add sound effects for an illustration.'' Jigsaw then asks GPT to use a chain of models to accomplish the task via the following prompt:


\begin{lstlisting}
You are given a set of AI models to complete a user's task.
There are thirty-nine models: <1. text2text() has reasoning capability. 2. text2img() can generate an image from text. 3. text2video() can generate a video from text, ...>
You can only use the models given. You do not have to use all the models. You will answer in a JSON format. Here is an example answer: <example combination of puzzle pieces written in a JSON format for the frontend to parse>. Your task is to <user-specified task>.
\end{lstlisting}

Prior work has found that asking GPT to evaluate its own results can improve the correctness of its responses \cite{press2022measuring}. We thus ask GPT to evaluate its own answers based on four criteria via the following prompt:

\newpage

\begin{lstlisting}
<Prompt from the previous step>
<Answer from the previous step>
Here are four criteria that the answer needs to satisfy. If any criteria are not satisfied, please give me the corrected answer in JSON format.
1. Whether the user's task was understood and completed.
2. Whether no models outside of the provided ones were used.
3. Whether the output and input of each step can be connected.
4. Whether it follows the correct JSON format.
\end{lstlisting}

Jigsaw then passes the chain of puzzle pieces provided in JSON format to the frontend and adds them onto the Assembly Panel. The designer can make further edits to the chain, just like any manually-created chain.

\subsection{System walkthrough}
\label{section:system-walkthrough}

\begin{figure*}[tbp]
  \centering
  \includegraphics[width=15cm]{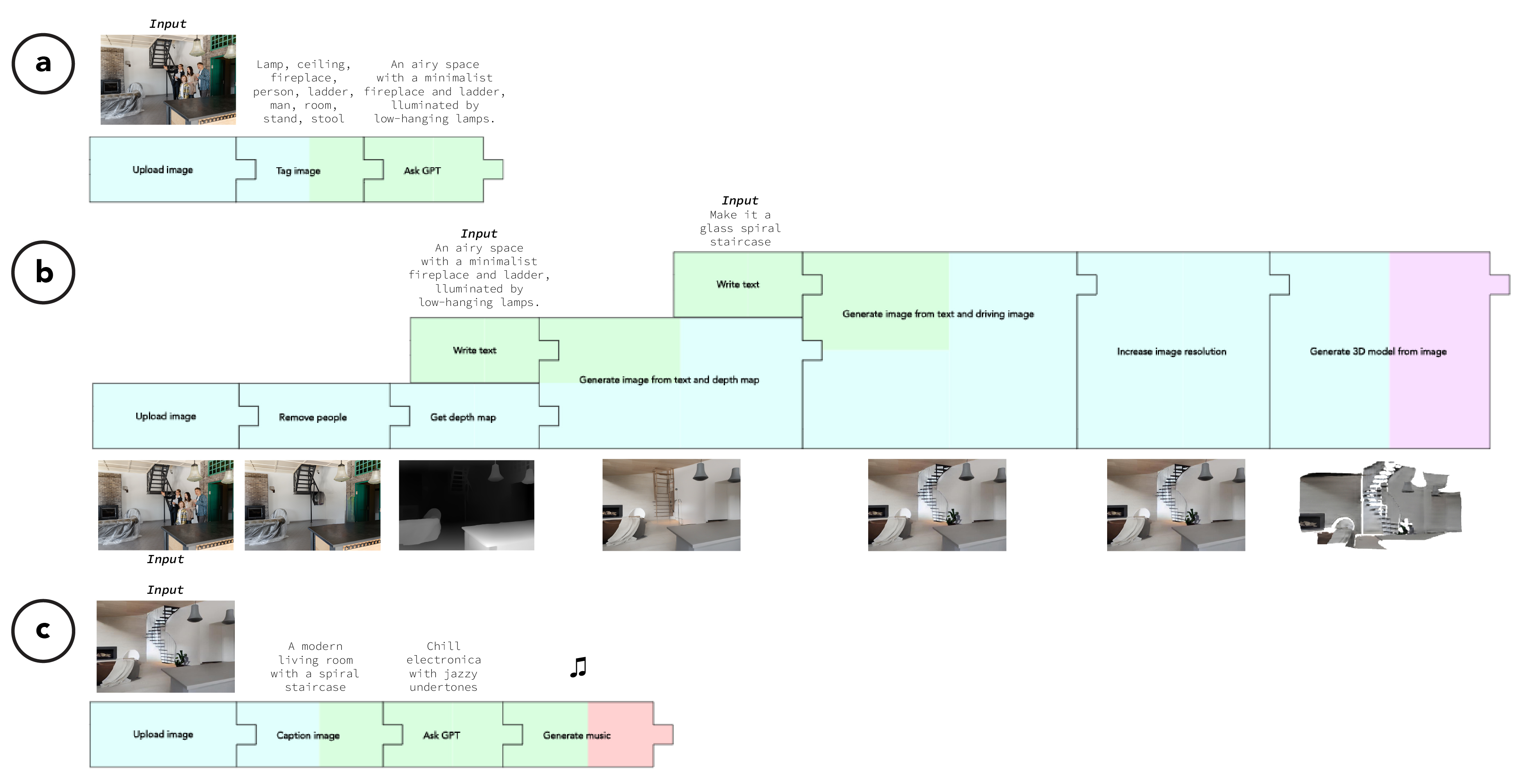}
  \caption{An example model mosaic for interior design. The designer can use Jigsaw to ideate a design concept from client material (a), design the 2D and 3D mockups (b), and add music to enhance the presentation (c).}
  \Description{An example model mosaic for interior design. The designer can use Jigsaw to ideate a design concept from client material (a), design the 2D and 3D mockups (b), and add music to enhance the presentation (c).}
  \label{fig:interior-design-model-mosaic}
\end{figure*}

We illustrate the interactions supported by Jigsaw using an interior design example. Figure \ref{fig:interior-design-model-mosaic} displays the final arrangement of puzzle pieces, referred to as the "model mosaic" in this paper for succinctness.

\subsubsection{Ideating a design concept from client material}

Zaha is an interior designer tasked with creating and presenting a redesigned interior for a client's new home. She received a photograph of interior from the client.

To begin, Zaha plans to create a design concept using the client's photo as a reference (Figure \ref{fig:interior-design-model-mosaic}a). She drags an \texttt{\fbox{\textcolor{blue}{Upload image}}} puzzle piece from the Catalog Panel onto the Assembly Panel, and uploads the client's photo in the Input Panel.
Zaha first identifies existing features in the client's home, such as built-in structures, furniture, and lights.
Thus, Zaha uses the semantic search bar in the Catalog Panel to find a puzzle piece that can \textit{"identify the objects inside the image"}. Jigsaw returns the \texttt{\fbox{\textcolor{green}{Tag \textcolor{blue}{image}}}} piece as the top result.
She adds \texttt{\fbox{\textcolor{green}{Tag} \textcolor{blue}{image}}} after \texttt{\fbox{\textcolor{blue}{Upload image}}} to identify the objects.
Zaha then uses the \texttt{\fbox{\textcolor{green}{Ask GPT}}} piece in \textit{ideation} mode, with \textit{"contemporary interior concept"} in the \textit{Task} box, to assist her in brainstorming a concept.
After clicking \textit{Run}, Jigsaw has suggested a design concept of \textit{"An airy space with a minimalist fireplace and ladder, illuminated by low-hanging lamps."}

\subsubsection{Designing the 2D and 3D mockups}

With a design concept in hand, Zaha proceeds to create the visual design (Figure \ref{fig:interior-design-model-mosaic}b). 
Zaha quickly duplicates the \texttt{\fbox{\textcolor{blue}{Upload image}}} piece to start a new chain, using the client's photo as a reference again.
She notices that the reference photo includes people, whom she wants to remove, and adds the \texttt{\fbox{\textcolor{blue}{Remove people}}} piece.
Zaha is interested in experimenting with AI image generation tools but acknowledges that the room's structure must remain intact for the design to be technically feasible.
\rev{She uses the semantic search bar in the Catalog Panel to find a puzzle piece that can help \textit{"preserve the structure of the room"}.
Jigsaw returns \texttt{\fbox{\textcolor{blue}{Get edge map}}} and \texttt{\fbox{\textcolor{blue}{Get depth map}}} as the top results.}
Zaha begins by testing the \texttt{\fbox{\textcolor{blue}{Get edge map}}} piece and adds the \texttt{\fbox{Generate \textcolor{blue}{image} from \textcolor{green}{text} and \textcolor{blue}{edge map}}} piece, which takes in both image and text inputs.
For text, she inputs the design concept suggested in the previous ideation chain. 

Zaha feels that the generated image fails to retain the desired room structure.
Recognizing this, she drags the pieces using edge maps into the trash bin.
She tries \texttt{\fbox{\textcolor{blue}{Get depth map}}} and
\texttt{\fbox{Generate}}
\texttt{\fbox{\textcolor{blue}{image} from \textcolor{green}{text} and \textcolor{blue}{depth map}}}
instead, which better preserves the room's structure. To try different design variations, Zaha inspects the parameters tooltip for the \texttt{\fbox{Generate \textcolor{blue}{image} from \textcolor{green}{text}}}
\texttt{\fbox{and \textcolor{blue}{depth map}}} model in the parameters sidebar. She discovers that she can tinker with the seed value to generate different variations.

Zaha is now satisfied with the redesign, except for the wooden ladder.
She believes replacing it with a spiral glass staircase would better fit the contemporary concept.
Thus, she searches for a puzzle piece that can modify an image using text instructions and finds the \texttt{\fbox{Generate \textcolor{blue}{image} from \textcolor{green}{text} and \textcolor{blue}{image}}} piece.
Zaha instructs it to \textit{"replace the wooden ladder with a glass spiral staircase."}
The newly generated redesign now features a contemporary glass spiral staircase.

Zaha would like to visualize a 3D mockup. She discovers the \texttt{\fbox{Generate \textcolor{indigo}{3D model} from \textcolor{blue}{image}}} piece and attaches this to the chain, but finds that the generated results are low resolution. Thus, Zaha searches the \textit{Image} section of the Catalog Panel for a piece that can help enhance the image. She finds the \texttt{\fbox{\textcolor{blue}{Increase image}}
\texttt{\fbox{resolution}}} piece.

\subsubsection{Presenting the design to the client}

To communicate the contemporary design concept, Zaha would like to incorporate a musical background to complement the design aesthetics.
To achieve this, Zaha asks the Assembly Assistant to \textit{"help add music based on the image"}. 
The Assembly Assistant suggests the following chain of puzzle pieces: \texttt{\fbox{\textcolor{green}{Caption} \textcolor{blue}{image}}} to understand the image, \texttt{\fbox{\textcolor{green}{Ask GPT}}} with \textit{ideation} mode to brainstorm a fitting music description, and \texttt{\fbox{\textcolor{green}{Generate} \textcolor{red}{music}}} to generate the music (Figure \ref{fig:interior-design-model-mosaic}c). This outcome is a chill electronic music piece.

\section{User Study}

We conducted a user study to understand how Jigsaw could address designers' pain points in working with multiple foundation models, its potential to be integrated into design workflows, and identify improvement areas.

\subsection{Participants and Procedure}

We invited the ten designers from our formative interviews to participate in a one-hour remote user study. They were not exposed to Jigsaw's system or concept prior to the user study. Participants accessed Jigsaw through a web browser, shared their screen, and verbally explained what they were doing and thinking (think-aloud).

\subsubsection{Introduction (10 minutes)}
Participants provided informed consent, and then received an introduction to Jigsaw's components, as described in Section \ref{section:implementation}.

\subsubsection{Reproduction Task (15 minutes)}
Participants were asked to reproduce the interior design model mosaic described in Section \ref{section:system-walkthrough}. Participants used the starter image shown in Figure \ref{fig:interior-design-model-mosaic}a and a detailed design brief of the various steps they would need to create (see Section \ref{section:system-walkthrough}).

\subsubsection{Free Creation Task (20 minutes)}
Participants were asked to freely explore Jigsaw and create their own model mosaics. We encouraged participants to build workflows beyond a simple chain and try out puzzle pieces involving multiple modalities.

\subsubsection{Post-Study Interview (15 minutes)}
After the creation activities, we conducted a semi-structured interview asking about participants' experiences using Jigsaw, whether they could see Jigsaw being integrated into their design workflow, and to identify areas for improving the system.

\subsection{Results, Discussion, and Future Work}

\begin{figure*}[tbp]
  \centering
  \includegraphics[width=9cm]{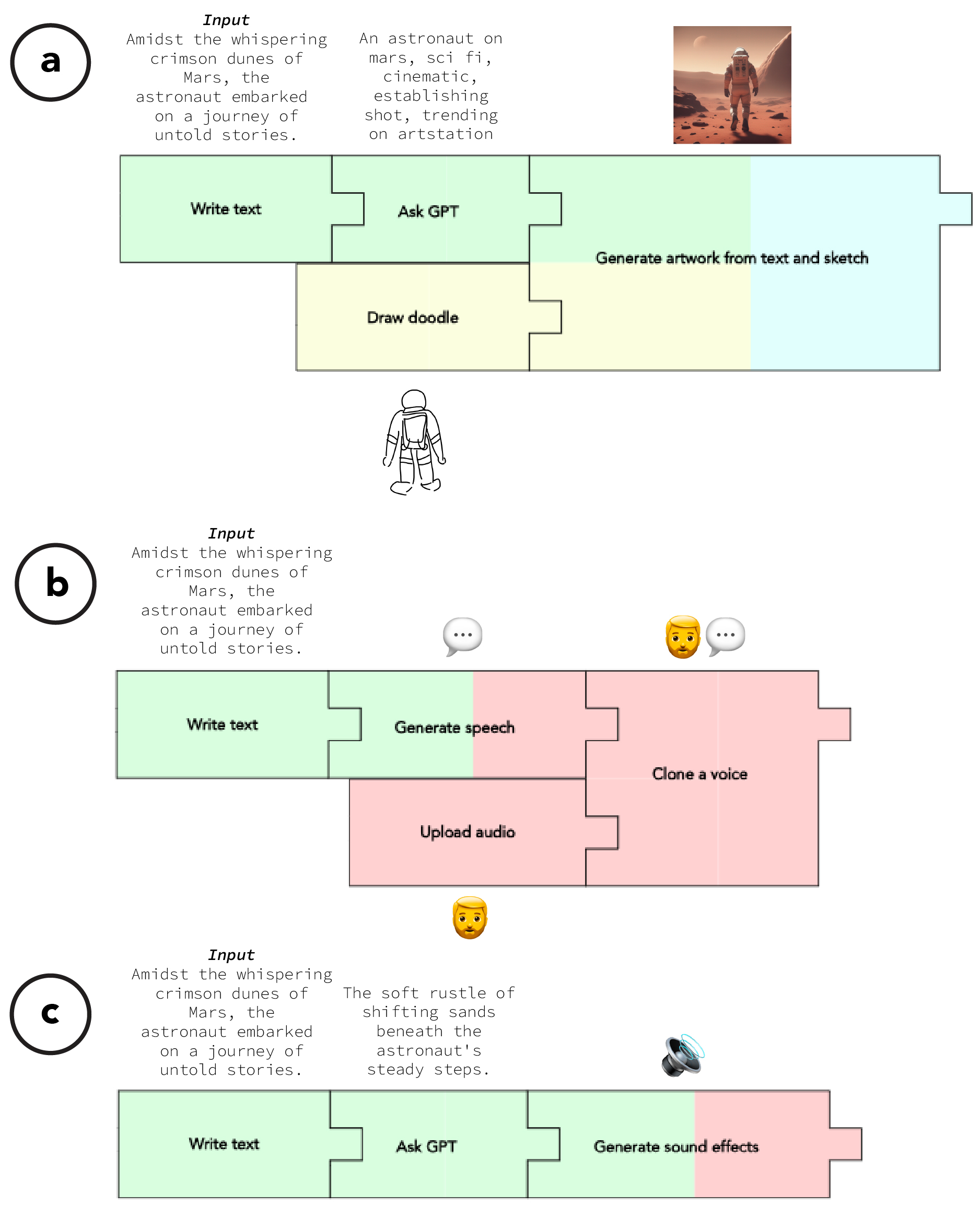}
  \caption{Model mosaic by P2, an illustrator, to create an audio-visual story. P2 uses Jigsaw to create an illustration based on a text description and a reference sketch (a), generate narrations through a cloned voice (b), and generate accompanying sound effects (c).}
  \Description{Model mosaic by P2, an illustrator, to create an audio-visual story. P2 uses Jigsaw to create an illustration based on a text description and a reference sketch (a), generate narrations through a cloned voice (b), and generate accompanying sound effects (c).}
  \label{fig:example-audio-visual-story}
\end{figure*}

\begin{figure*}[tbp]
  \centering
  \includegraphics[width=15cm]{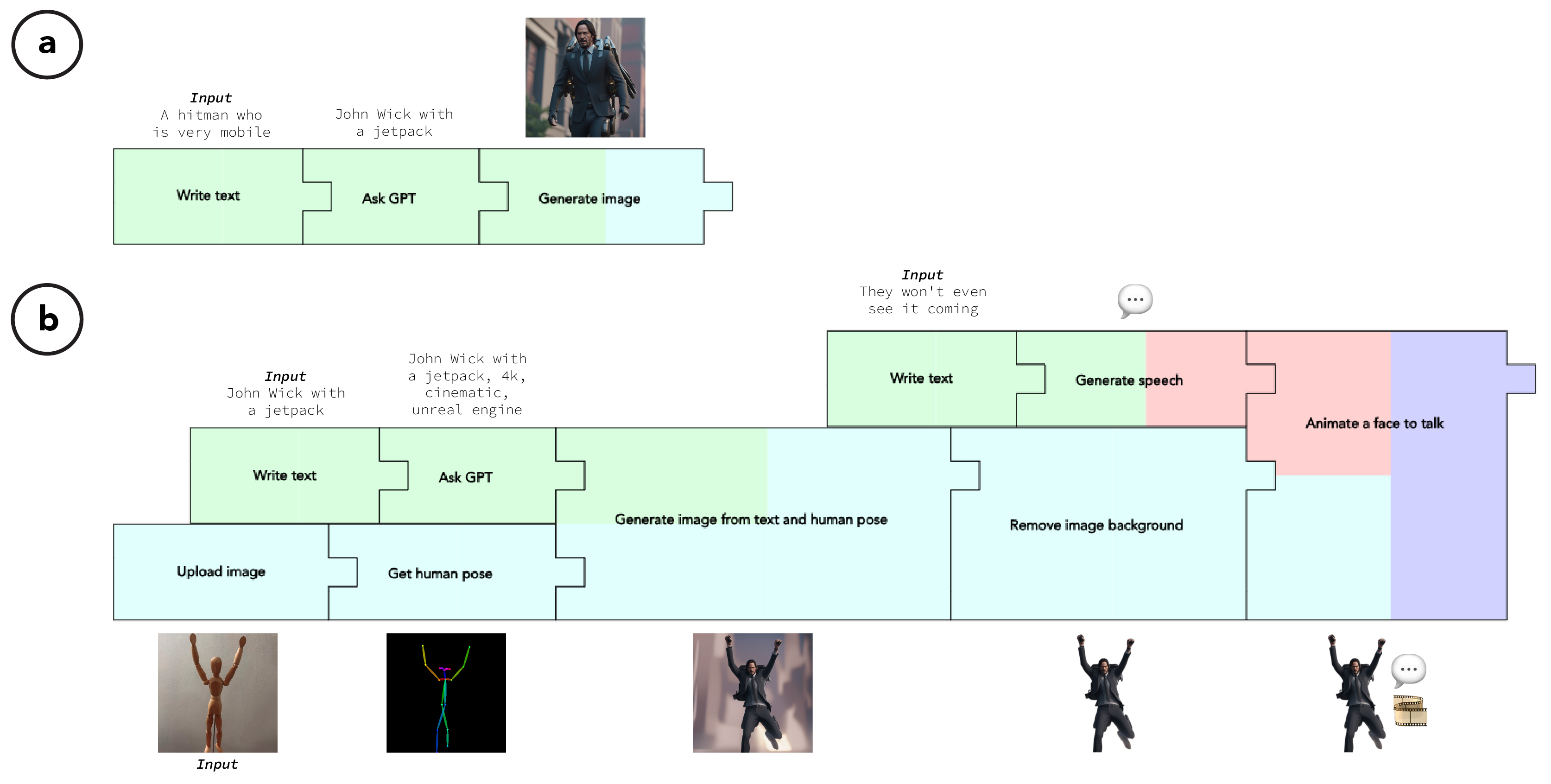}
  \caption{Model mosaic by P10, a game designer, to create a video game character. P10 uses Jigsaw to create a character concept and preview the character's visuals (a). P10 then specifies a pose for the character and generates the character's visuals in the specified pose (b). P10 then generates a line for the character to say and animates the character to deliver the line.}
  \Description{Model mosaic by P10, a game designer, to create a video game character. P10 uses Jigsaw to create a character concept and preview the character's visuals (a). P10 then used specifies a pose for the character and generates the character's visuals in the specified pose (b). P10 then generates a line for the character to say and animates the character to deliver the line.}
  \label{fig:example-video-game-character}
\end{figure*}

All participants completed the reproduction and free creation tasks. Jigsaw appears to help designers discover and prototype new creative workflows.
Designers suggested different future improvements.

\subsubsection{Helping designers discover and utilize AI capabilities}

Participants located AI capabilities via the Catalog Panel's semantic search bar or by filtering pieces by modality. 
Many participants mentioned that they were able to \textit{"discover \underline{new} AI abilities [they were] previously unaware of (P9)"}.
For example, P2, an illustrator, discovered the capabilities of ControlNet \cite{zhang2023adding}, a model that allows users to add additional control to a text-to-image model, such as a guiding sketch.
Figure \ref{fig:example-audio-visual-story} shows the model mosaic created by P2, who used Jigsaw to create an audio-visual story.
In Figure \ref{fig:example-audio-visual-story}a, she created visuals for her story. Instead of using a text-to-image model, she discovered and used ControlNet to generate images based on a starting sketch.
In total, participants used 8 model puzzle pieces on average ($\mu$=7.9, $\sigma$=1.29), and all participants explored beyond well-known models such as GPT and Stable Diffusion (see Appendix \ref{appendix:user-chains} for details).
As the number of foundation models continues to increase, we plan to expand our set of puzzle pieces for designers over time.

Participants commented that tooltips \textit{"gave [them] a solid understanding of the capabilities of each of the puzzle pieces, like what can be expected as output and what types of inputs are suitable (P1)"}. In addition, participants expressed that the assisted prompting mechanism offered by the \textit{translation} glue piece \textit{"allowed [them] to achieve satisfying results without the need to laboriously rephrase and tweak prompts (P2)"}: \textit{"Now, it's like I speak the AI's language. (P5)"}.

\subsubsection{Supporting intuitive prototyping}

Participants expressed that \textit{"[they] enjoyed the idea of building with AI visually with tangible puzzle pieces (P10)"} and found it \textit{"easy to pick up and start designing (P3)"}. 
In particular, participants appreciated the error-proof design: \textit{"Knowing which puzzle pieces can be connected expedites my prototyping. I see the same colors and receive snapping feedback. I don't spend time building a workflow and then compile it to find compatibility errors (P3)"}. A contribution of this research is showing how the design benefits of block-based VPIs, commonly catered towards novice programming learners \cite{resnick2009scratch, fraser2013blockly, weintrop2015block}, can be effectively applied to the realm of design prototyping for non-technical designers to work with AI capabilities. An interesting extension of Jigsaw could be the implementation of a "tutorial mode" to teach novice designers.
The system would disassemble a model mosaic created by an experienced designer into pieces, allowing a novice designer to recreate it and learn from the experienced designer’s design process.

Furthermore, participants mentioned that the ability to near-instantaneously see intermediate outputs in a chain \textit{"helped [them] to quickly test out ideas and make adjustments to individual steps as needed (P5)"}.
This aligns with findings from prior research in interactive program debugging tools \cite{karsai1995configurable, schoop2021umlaut, jiang2023log}.

\subsubsection{Serving as a brainstorming and documentation canvas}

We observed participants making creative uses of the Assembly Panel, including using it to test different partial workflows before combining them into more complete workflows and using it to document their design explorations.
Participants expressed that the Assembly Panel provides \textit{"a playground to be messy and experimental (P3)"} and \textit{"makes it easy to track the evolution of an idea (P4)."}
Moreover, participants commented that the \textit{ideation} glue piece was \textit{"helpful for brainstorming concepts at the beginning [of the design process] (P2)"}.
We observed that designers occasionally passed the outputs of the \textit{ideation} glue directly into a generation model, as shown in Figure \ref{fig:example-audio-visual-story}c. In other instances, designers maintained a shorter chain solely for concept generation. This is shown in the model mosaic created by P10, a game designer, in Figure \ref{fig:example-video-game-character}, who created a video game character. He primarily used the short chain in Figure \ref{fig:example-video-game-character}a to generate concepts for his character. We observed that designers frequently created multiple chains to organize different stages of their design process.
Participants noted that since the canvas documents their creative process, it could serve as \textit{"a boundary object for sharing and explaining ideas to clients (P5)"}.

Moreover, participants commented that the Assembly Assistant was useful in \textit{"generating an initial configuration of puzzle pieces to start working with (P1)"}. This aids in combating the \textit{"blank canvas syndrome (P6)"}, a common occurrence at the onset of a creative activity \cite{joyce2009blank}.
In Figure \ref{fig:example-video-game-character}b, P10 wanted his video game character to look like Superman flying.
However, he initially struggled to come up with a method to accomplish this, so he sought assistance from the Assembly Assistant. The Assembly Assistant recommended a workflow of a reference pose image, a pose extraction model, and a ControlNet model that can be guided by pose.
Given this workflow, P10 used a wooden mannequin to specify the pose for his character.

\rev{\section{Limitations and Future Work}}
\rev{There are several avenues for improvement that we plan to address for future work.
First, we currently implemented one AI model for each design task (e.g., Stable Diffusion for text-to-image). We plan to support the capability of switching between multiple alternative models. We will provide information on the tradeoffs between them (e.g., speed vs. quality) for both the user and as context for Jigsaw's subcomponents (i.e., semantic search and Assembly Assistant), facilitate easy side-by-side comparison, and allow users to filter models by certain criteria (e.g., text-to-image models with a typical runtime of under 10 seconds).
Second, we are interested in expanding Jigsaw to let designers define custom puzzle pieces, as suggested by P10, and logic operators such as if/else statements and loops, common in other VPIs \cite{max}.
Third, we currently use LLMs as glue, using the text modality for intermediate reasoning (Section \ref{section:glue}). We anticipate extending the glue piece to incorporate newer research on multimodal LLMs (MLLMs), such as GPT-4V \cite{gpt4v} and LLaVA \cite{liu2023visual}, to add information from additional modalities.
Fourth,  we plan to expand the Input and Output panels to handle real-time video and audio streams, as suggested by P9.
Finally, participants noted that the Assembly Assistant was less robust to ambiguous tasks or tasks that require very complicated mosaics.
As the Assembly Assistant uses GPT, we anticipate improvements to the Assembly Assistant as stronger versions of GPT are released.
This is also a common challenge recognized by recent machine learning works that also aim to \textit{automatically} combine expert models to solve complex AI tasks \cite{shen2023hugginggpt, wu2023visual, huang2023audiogpt}.
A path forward, as suggested by P7, could be to improve the Assembly Assistant to support \textit{back-and-forth interactions} with the designer, becoming a design co-pilot that assists designers in creating complex workflows.
As more designers use Jigsaw to create model mosaics, we plan to compile them into a template gallery that other designers can modify for their own use cases, as suggested by P8. 
We believe that the accumulated design templates can serve as a search space for the Assembly Assistant, enhancing its capabilities as a design search engine.}

\section{Conclusion}
This research identifies the challenges designers face when using AI foundation models to support their work.
The research prototype, Jigsaw, uses a puzzle piece metaphor to represent foundation models and allows the combination of models by assembling compatible pieces.
Feedback from designers using Jigsaw demonstrated that designers discovered new AI capabilities, combined multiple AI capabilities across various modalities, and flexibly explored, prototyped, and documented AI-enabled design workflows.
We are interested in extending Jigsaw with more capabilities and hope that this research can help inform future research on block-based prototyping systems for prototyping with AI foundation models.



\begin{acks}
We would like to thank friends in the Augmented Design Capability Studio for providing valuable feedback on iterations of our system and paper write-up, and the anonymous reviewers for their comments.
\end{acks}

\bibliographystyle{ACM-Reference-Format}
\bibliography{base}


\begin{thebibliography}{58}


\ifx \showCODEN    \undefined \def \showCODEN     #1{\unskip}     \fi
\ifx \showDOI      \undefined \def \showDOI       #1{#1}\fi
\ifx \showISBNx    \undefined \def \showISBNx     #1{\unskip}     \fi
\ifx \showISBNxiii \undefined \def \showISBNxiii  #1{\unskip}     \fi
\ifx \showISSN     \undefined \def \showISSN      #1{\unskip}     \fi
\ifx \showLCCN     \undefined \def \showLCCN      #1{\unskip}     \fi
\ifx \shownote     \undefined \def \shownote      #1{#1}          \fi
\ifx \showarticletitle \undefined \def \showarticletitle #1{#1}   \fi
\ifx \showURL      \undefined \def \showURL       {\relax}        \fi
\providecommand\bibfield[2]{#2}
\providecommand\bibinfo[2]{#2}
\providecommand\natexlab[1]{#1}
\providecommand\showeprint[2][]{arXiv:#2}

\bibitem[com(2013)]%
        {comfyui}
 \bibinfo{year}{2013}\natexlab{}.
\newblock \bibinfo{title}{ComfyUI}.
\newblock \bibinfo{howpublished}{\url{https://github.com/comfyanonymous/ComfyUI}}.
\newblock


\bibitem[flo(2013)]%
        {flowiseai}
 \bibinfo{year}{2013}\natexlab{}.
\newblock \bibinfo{title}{FlowiseAI}.
\newblock \bibinfo{howpublished}{\url{https://github.com/FlowiseAI/Flowise}}.
\newblock


\bibitem[lan(2013)]%
        {langflow}
 \bibinfo{year}{2013}\natexlab{}.
\newblock \bibinfo{title}{Langflow}.
\newblock \bibinfo{howpublished}{\url{https://github.com/logspace-ai/langflow}}.
\newblock


\bibitem[run(2023)]%
        {runwayml}
 \bibinfo{year}{2023}\natexlab{}.
\newblock \bibinfo{booktitle}{\emph{AI Magic Tools - Runway}}.
\newblock
\urldef\tempurl%
\url{https://runwayml.com/ai-magic-tools/}
\showURL{%
Retrieved December 10, 2023 from \tempurl}


\bibitem[cha(2023)]%
        {chatgpt}
 \bibinfo{year}{2023}\natexlab{}.
\newblock \bibinfo{booktitle}{\emph{ChatGPT}}.
\newblock
\urldef\tempurl%
\url{https://chat.openai.com/}
\showURL{%
Retrieved August 15, 2023 from \tempurl}


\bibitem[gpt(2023)]%
        {gpt4v}
 \bibinfo{year}{2023}\natexlab{}.
\newblock \bibinfo{booktitle}{\emph{GPT-4V(ision) System Card}}.
\newblock
\urldef\tempurl%
\url{https://cdn.openai.com/papers/GPTV_System_Card.pdf}
\showURL{%
Retrieved December 10, 2023 from \tempurl}


\bibitem[mid(2023)]%
        {midjourney}
 \bibinfo{year}{2023}\natexlab{}.
\newblock \bibinfo{booktitle}{\emph{Midjourney}}.
\newblock
\urldef\tempurl%
\url{https://www.midjourney.com/}
\showURL{%
Retrieved August 15, 2023 from \tempurl}


\bibitem[fou(2023)]%
        {foundation-models}
 \bibinfo{year}{2023}\natexlab{}.
\newblock \bibinfo{booktitle}{\emph{Reflections on Foundation Models}}.
\newblock
\urldef\tempurl%
\url{https://hai.stanford.edu/news/reflections-foundation-models}
\showURL{%
Retrieved August 15, 2023 from \tempurl}


\bibitem[gra(2023)]%
        {grasshopper}
 \bibinfo{year}{2023}\natexlab{}.
\newblock \bibinfo{booktitle}{\emph{Rhino - Grasshopper - New in Rhino 6}}.
\newblock
\urldef\tempurl%
\url{https://www.rhino3d.com/6/new/grasshopper/}
\showURL{%
Retrieved August 15, 2023 from \tempurl}


\bibitem[tra(2023)]%
        {transfer-learning}
 \bibinfo{year}{2023}\natexlab{}.
\newblock \bibinfo{booktitle}{\emph{Transfer learning and fine-tuning}}.
\newblock
\urldef\tempurl%
\url{https://www.tensorflow.org/tutorials/images/transfer_learning}
\showURL{%
Retrieved August 15, 2023 from \tempurl}


\bibitem[uni(2023)]%
        {unity}
 \bibinfo{year}{2023}\natexlab{}.
\newblock \bibinfo{booktitle}{\emph{Unity Visual Scripting}}.
\newblock
\urldef\tempurl%
\url{https://unity.com/features/unity-visual-scripting}
\showURL{%
Retrieved August 15, 2023 from \tempurl}


\bibitem[max(2023)]%
        {max}
 \bibinfo{year}{2023}\natexlab{}.
\newblock \bibinfo{booktitle}{\emph{What is Max? | Cycling '74}}.
\newblock
\urldef\tempurl%
\url{https://cycling74.com/products/max}
\showURL{%
Retrieved August 15, 2023 from \tempurl}


\bibitem[Bommasani et~al\mbox{.}(2021)]%
        {bommasani2021opportunities}
\bibfield{author}{\bibinfo{person}{Rishi Bommasani}, \bibinfo{person}{Drew~A Hudson}, \bibinfo{person}{Ehsan Adeli}, \bibinfo{person}{Russ Altman}, \bibinfo{person}{Simran Arora}, \bibinfo{person}{Sydney von Arx}, \bibinfo{person}{Michael~S Bernstein}, \bibinfo{person}{Jeannette Bohg}, \bibinfo{person}{Antoine Bosselut}, \bibinfo{person}{Emma Brunskill}, {et~al\mbox{.}}} \bibinfo{year}{2021}\natexlab{}.
\newblock \showarticletitle{On the opportunities and risks of foundation models}.
\newblock \bibinfo{journal}{\emph{arXiv preprint arXiv:2108.07258}} (\bibinfo{year}{2021}).
\newblock


\bibitem[Brown et~al\mbox{.}(2020)]%
        {brown2020language}
\bibfield{author}{\bibinfo{person}{Tom Brown}, \bibinfo{person}{Benjamin Mann}, \bibinfo{person}{Nick Ryder}, \bibinfo{person}{Melanie Subbiah}, \bibinfo{person}{Jared~D Kaplan}, \bibinfo{person}{Prafulla Dhariwal}, \bibinfo{person}{Arvind Neelakantan}, \bibinfo{person}{Pranav Shyam}, \bibinfo{person}{Girish Sastry}, \bibinfo{person}{Amanda Askell}, {et~al\mbox{.}}} \bibinfo{year}{2020}\natexlab{}.
\newblock \showarticletitle{Language models are few-shot learners}.
\newblock \bibinfo{journal}{\emph{Advances in neural information processing systems}}  \bibinfo{volume}{33} (\bibinfo{year}{2020}), \bibinfo{pages}{1877--1901}.
\newblock


\bibitem[Carney et~al\mbox{.}(2020)]%
        {carney2020teachable}
\bibfield{author}{\bibinfo{person}{Michelle Carney}, \bibinfo{person}{Barron Webster}, \bibinfo{person}{Irene Alvarado}, \bibinfo{person}{Kyle Phillips}, \bibinfo{person}{Noura Howell}, \bibinfo{person}{Jordan Griffith}, \bibinfo{person}{Jonas Jongejan}, \bibinfo{person}{Amit Pitaru}, {and} \bibinfo{person}{Alexander Chen}.} \bibinfo{year}{2020}\natexlab{}.
\newblock \showarticletitle{Teachable machine: Approachable Web-based tool for exploring machine learning classification}. In \bibinfo{booktitle}{\emph{Extended abstracts of the 2020 CHI conference on human factors in computing systems}}. \bibinfo{pages}{1--8}.
\newblock


\bibitem[Chiou et~al\mbox{.}(2023)]%
        {chiou2023designing}
\bibfield{author}{\bibinfo{person}{Li-Yuan Chiou}, \bibinfo{person}{Peng-Kai Hung}, \bibinfo{person}{Rung-Huei Liang}, {and} \bibinfo{person}{Chun-Teng Wang}.} \bibinfo{year}{2023}\natexlab{}.
\newblock \showarticletitle{Designing with AI: An Exploration of Co-Ideation with Image Generators}. In \bibinfo{booktitle}{\emph{Proceedings of the 2023 ACM Designing Interactive Systems Conference}}. \bibinfo{pages}{1941--1954}.
\newblock


\bibitem[Cox et~al\mbox{.}(1989)]%
        {cox1989prograph}
\bibfield{author}{\bibinfo{person}{Philip~T Cox}, \bibinfo{person}{FR Giles}, {and} \bibinfo{person}{Tomasz Pietrzykowski}.} \bibinfo{year}{1989}\natexlab{}.
\newblock \showarticletitle{Prograph: a step towards liberating programming from textual conditioning}. In \bibinfo{booktitle}{\emph{1989 IEEE Workshop on Visual languages}}. IEEE Computer Society, \bibinfo{pages}{150--151}.
\newblock


\bibitem[Du et~al\mbox{.}(2023)]%
        {du2023rapsai}
\bibfield{author}{\bibinfo{person}{Ruofei Du}, \bibinfo{person}{Na Li}, \bibinfo{person}{Jing Jin}, \bibinfo{person}{Michelle Carney}, \bibinfo{person}{Scott Miles}, \bibinfo{person}{Maria Kleiner}, \bibinfo{person}{Xiuxiu Yuan}, \bibinfo{person}{Yinda Zhang}, \bibinfo{person}{Anuva Kulkarni}, \bibinfo{person}{Xingyu Liu}, {et~al\mbox{.}}} \bibinfo{year}{2023}\natexlab{}.
\newblock \showarticletitle{Rapsai: Accelerating Machine Learning Prototyping of Multimedia Applications through Visual Programming}. In \bibinfo{booktitle}{\emph{Proceedings of the 2023 CHI Conference on Human Factors in Computing Systems}}. \bibinfo{pages}{1--23}.
\newblock


\bibitem[Fraser et~al\mbox{.}(2013)]%
        {fraser2013blockly}
\bibfield{author}{\bibinfo{person}{Neil Fraser} {et~al\mbox{.}}} \bibinfo{year}{2013}\natexlab{}.
\newblock \showarticletitle{Blockly: A visual programming editor}.
\newblock \bibinfo{journal}{\emph{URL: https://code. google. com/p/blockly}}  \bibinfo{volume}{42} (\bibinfo{year}{2013}).
\newblock


\bibitem[Gmeiner et~al\mbox{.}(2023)]%
        {gmeiner2023exploring}
\bibfield{author}{\bibinfo{person}{Frederic Gmeiner}, \bibinfo{person}{Humphrey Yang}, \bibinfo{person}{Lining Yao}, \bibinfo{person}{Kenneth Holstein}, {and} \bibinfo{person}{Nikolas Martelaro}.} \bibinfo{year}{2023}\natexlab{}.
\newblock \showarticletitle{Exploring Challenges and Opportunities to Support Designers in Learning to Co-create with AI-based Manufacturing Design Tools}. In \bibinfo{booktitle}{\emph{Proceedings of the 2023 CHI Conference on Human Factors in Computing Systems}}. \bibinfo{pages}{1--20}.
\newblock


\bibitem[Huang et~al\mbox{.}(2023)]%
        {huang2023audiogpt}
\bibfield{author}{\bibinfo{person}{Rongjie Huang}, \bibinfo{person}{Mingze Li}, \bibinfo{person}{Dongchao Yang}, \bibinfo{person}{Jiatong Shi}, \bibinfo{person}{Xuankai Chang}, \bibinfo{person}{Zhenhui Ye}, \bibinfo{person}{Yuning Wu}, \bibinfo{person}{Zhiqing Hong}, \bibinfo{person}{Jiawei Huang}, \bibinfo{person}{Jinglin Liu}, {et~al\mbox{.}}} \bibinfo{year}{2023}\natexlab{}.
\newblock \showarticletitle{Audiogpt: Understanding and generating speech, music, sound, and talking head}.
\newblock \bibinfo{journal}{\emph{arXiv preprint arXiv:2304.12995}} (\bibinfo{year}{2023}).
\newblock


\bibitem[Jiang et~al\mbox{.}(2023)]%
        {jiang2023log}
\bibfield{author}{\bibinfo{person}{Peiling Jiang}, \bibinfo{person}{Fuling Sun}, {and} \bibinfo{person}{Haijun Xia}.} \bibinfo{year}{2023}\natexlab{}.
\newblock \showarticletitle{Log-it: Supporting Programming with Interactive, Contextual, Structured, and Visual Logs}. In \bibinfo{booktitle}{\emph{Proceedings of the 2023 CHI Conference on Human Factors in Computing Systems}}. \bibinfo{pages}{1--16}.
\newblock


\bibitem[Joyce(2009)]%
        {joyce2009blank}
\bibfield{author}{\bibinfo{person}{Caneel~K Joyce}.} \bibinfo{year}{2009}\natexlab{}.
\newblock \bibinfo{booktitle}{\emph{The blank page: Effects of constraint on creativity}}.
\newblock \bibinfo{publisher}{University of California, Berkeley}.
\newblock


\bibitem[Jun and Nichol(2023)]%
        {jun2023shap}
\bibfield{author}{\bibinfo{person}{Heewoo Jun} {and} \bibinfo{person}{Alex Nichol}.} \bibinfo{year}{2023}\natexlab{}.
\newblock \showarticletitle{Shap-e: Generating conditional 3d implicit functions}.
\newblock \bibinfo{journal}{\emph{arXiv preprint arXiv:2305.02463}} (\bibinfo{year}{2023}).
\newblock


\bibitem[Karsai(1995)]%
        {karsai1995configurable}
\bibfield{author}{\bibinfo{person}{G{\'a}bor Karsai}.} \bibinfo{year}{1995}\natexlab{}.
\newblock \showarticletitle{A configurable visual programming environment: A tool for domain-specific programming}.
\newblock \bibinfo{journal}{\emph{Computer}} \bibinfo{volume}{28}, \bibinfo{number}{3} (\bibinfo{year}{1995}), \bibinfo{pages}{36--44}.
\newblock


\bibitem[Kirillov et~al\mbox{.}(2023)]%
        {kirillov2023segment}
\bibfield{author}{\bibinfo{person}{Alexander Kirillov}, \bibinfo{person}{Eric Mintun}, \bibinfo{person}{Nikhila Ravi}, \bibinfo{person}{Hanzi Mao}, \bibinfo{person}{Chloe Rolland}, \bibinfo{person}{Laura Gustafson}, \bibinfo{person}{Tete Xiao}, \bibinfo{person}{Spencer Whitehead}, \bibinfo{person}{Alexander~C Berg}, \bibinfo{person}{Wan-Yen Lo}, {et~al\mbox{.}}} \bibinfo{year}{2023}\natexlab{}.
\newblock \showarticletitle{Segment anything}.
\newblock \bibinfo{journal}{\emph{arXiv preprint arXiv:2304.02643}} (\bibinfo{year}{2023}).
\newblock


\bibitem[Kodosky(2020)]%
        {kodosky2020labview}
\bibfield{author}{\bibinfo{person}{Jeffrey Kodosky}.} \bibinfo{year}{2020}\natexlab{}.
\newblock \showarticletitle{LabVIEW}.
\newblock \bibinfo{journal}{\emph{Proceedings of the ACM on Programming Languages}} \bibinfo{volume}{4}, \bibinfo{number}{HOPL} (\bibinfo{year}{2020}), \bibinfo{pages}{1--54}.
\newblock


\bibitem[Kreuk et~al\mbox{.}(2022)]%
        {kreuk2022audiogen}
\bibfield{author}{\bibinfo{person}{Felix Kreuk}, \bibinfo{person}{Gabriel Synnaeve}, \bibinfo{person}{Adam Polyak}, \bibinfo{person}{Uriel Singer}, \bibinfo{person}{Alexandre D{\'e}fossez}, \bibinfo{person}{Jade Copet}, \bibinfo{person}{Devi Parikh}, \bibinfo{person}{Yaniv Taigman}, {and} \bibinfo{person}{Yossi Adi}.} \bibinfo{year}{2022}\natexlab{}.
\newblock \showarticletitle{Audiogen: Textually guided audio generation}.
\newblock \bibinfo{journal}{\emph{arXiv preprint arXiv:2209.15352}} (\bibinfo{year}{2022}).
\newblock


\bibitem[Liu et~al\mbox{.}(2023)]%
        {liu2023visual}
\bibfield{author}{\bibinfo{person}{Haotian Liu}, \bibinfo{person}{Chunyuan Li}, \bibinfo{person}{Qingyang Wu}, {and} \bibinfo{person}{Yong~Jae Lee}.} \bibinfo{year}{2023}\natexlab{}.
\newblock \showarticletitle{Visual instruction tuning}.
\newblock \bibinfo{journal}{\emph{arXiv preprint arXiv:2304.08485}} (\bibinfo{year}{2023}).
\newblock


\bibitem[Liu and Chilton(2022)]%
        {liu2022design}
\bibfield{author}{\bibinfo{person}{Vivian Liu} {and} \bibinfo{person}{Lydia~B Chilton}.} \bibinfo{year}{2022}\natexlab{}.
\newblock \showarticletitle{Design guidelines for prompt engineering text-to-image generative models}. In \bibinfo{booktitle}{\emph{Proceedings of the 2022 CHI Conference on Human Factors in Computing Systems}}. \bibinfo{pages}{1--23}.
\newblock


\bibitem[Myers(1990)]%
        {myers1990taxonomies}
\bibfield{author}{\bibinfo{person}{Brad~A Myers}.} \bibinfo{year}{1990}\natexlab{}.
\newblock \showarticletitle{Taxonomies of visual programming and program visualization}.
\newblock \bibinfo{journal}{\emph{Journal of Visual Languages \& Computing}} \bibinfo{volume}{1}, \bibinfo{number}{1} (\bibinfo{year}{1990}), \bibinfo{pages}{97--123}.
\newblock


\bibitem[Press et~al\mbox{.}(2022)]%
        {press2022measuring}
\bibfield{author}{\bibinfo{person}{Ofir Press}, \bibinfo{person}{Muru Zhang}, \bibinfo{person}{Sewon Min}, \bibinfo{person}{Ludwig Schmidt}, \bibinfo{person}{Noah~A Smith}, {and} \bibinfo{person}{Mike Lewis}.} \bibinfo{year}{2022}\natexlab{}.
\newblock \showarticletitle{Measuring and narrowing the compositionality gap in language models}.
\newblock \bibinfo{journal}{\emph{arXiv preprint arXiv:2210.03350}} (\bibinfo{year}{2022}).
\newblock


\bibitem[Radford et~al\mbox{.}(2021)]%
        {radford2021learning}
\bibfield{author}{\bibinfo{person}{Alec Radford}, \bibinfo{person}{Jong~Wook Kim}, \bibinfo{person}{Chris Hallacy}, \bibinfo{person}{Aditya Ramesh}, \bibinfo{person}{Gabriel Goh}, \bibinfo{person}{Sandhini Agarwal}, \bibinfo{person}{Girish Sastry}, \bibinfo{person}{Amanda Askell}, \bibinfo{person}{Pamela Mishkin}, \bibinfo{person}{Jack Clark}, {et~al\mbox{.}}} \bibinfo{year}{2021}\natexlab{}.
\newblock \showarticletitle{Learning transferable visual models from natural language supervision}. In \bibinfo{booktitle}{\emph{International conference on machine learning}}. PMLR, \bibinfo{pages}{8748--8763}.
\newblock


\bibitem[Ramesh et~al\mbox{.}(2021)]%
        {ramesh2021zero}
\bibfield{author}{\bibinfo{person}{Aditya Ramesh}, \bibinfo{person}{Mikhail Pavlov}, \bibinfo{person}{Gabriel Goh}, \bibinfo{person}{Scott Gray}, \bibinfo{person}{Chelsea Voss}, \bibinfo{person}{Alec Radford}, \bibinfo{person}{Mark Chen}, {and} \bibinfo{person}{Ilya Sutskever}.} \bibinfo{year}{2021}\natexlab{}.
\newblock \showarticletitle{Zero-shot text-to-image generation}. In \bibinfo{booktitle}{\emph{International Conference on Machine Learning}}. PMLR, \bibinfo{pages}{8821--8831}.
\newblock


\bibitem[Resnick et~al\mbox{.}(2009)]%
        {resnick2009scratch}
\bibfield{author}{\bibinfo{person}{Mitchel Resnick}, \bibinfo{person}{John Maloney}, \bibinfo{person}{Andr{\'e}s Monroy-Hern{\'a}ndez}, \bibinfo{person}{Natalie Rusk}, \bibinfo{person}{Evelyn Eastmond}, \bibinfo{person}{Karen Brennan}, \bibinfo{person}{Amon Millner}, \bibinfo{person}{Eric Rosenbaum}, \bibinfo{person}{Jay Silver}, \bibinfo{person}{Brian Silverman}, {et~al\mbox{.}}} \bibinfo{year}{2009}\natexlab{}.
\newblock \showarticletitle{Scratch: programming for all}.
\newblock \bibinfo{journal}{\emph{Commun. ACM}} \bibinfo{volume}{52}, \bibinfo{number}{11} (\bibinfo{year}{2009}), \bibinfo{pages}{60--67}.
\newblock


\bibitem[Rombach et~al\mbox{.}(2022)]%
        {rombach2022high}
\bibfield{author}{\bibinfo{person}{Robin Rombach}, \bibinfo{person}{Andreas Blattmann}, \bibinfo{person}{Dominik Lorenz}, \bibinfo{person}{Patrick Esser}, {and} \bibinfo{person}{Bj{\"o}rn Ommer}.} \bibinfo{year}{2022}\natexlab{}.
\newblock \showarticletitle{High-resolution image synthesis with latent diffusion models}. In \bibinfo{booktitle}{\emph{Proceedings of the IEEE/CVF conference on computer vision and pattern recognition}}. \bibinfo{pages}{10684--10695}.
\newblock


\bibitem[Schoop et~al\mbox{.}(2021)]%
        {schoop2021umlaut}
\bibfield{author}{\bibinfo{person}{Eldon Schoop}, \bibinfo{person}{Forrest Huang}, {and} \bibinfo{person}{Bjoern Hartmann}.} \bibinfo{year}{2021}\natexlab{}.
\newblock \showarticletitle{Umlaut: Debugging deep learning programs using program structure and model behavior}. In \bibinfo{booktitle}{\emph{Proceedings of the 2021 CHI conference on human factors in computing systems}}. \bibinfo{pages}{1--16}.
\newblock


\bibitem[Shen et~al\mbox{.}(2023)]%
        {shen2023hugginggpt}
\bibfield{author}{\bibinfo{person}{Yongliang Shen}, \bibinfo{person}{Kaitao Song}, \bibinfo{person}{Xu Tan}, \bibinfo{person}{Dongsheng Li}, \bibinfo{person}{Weiming Lu}, {and} \bibinfo{person}{Yueting Zhuang}.} \bibinfo{year}{2023}\natexlab{}.
\newblock \showarticletitle{Hugginggpt: Solving ai tasks with chatgpt and its friends in huggingface}.
\newblock \bibinfo{journal}{\emph{arXiv preprint arXiv:2303.17580}} (\bibinfo{year}{2023}).
\newblock


\bibitem[Shi et~al\mbox{.}(2023)]%
        {shi2023understanding}
\bibfield{author}{\bibinfo{person}{Yang Shi}, \bibinfo{person}{Tian Gao}, \bibinfo{person}{Xiaohan Jiao}, {and} \bibinfo{person}{Nan Cao}.} \bibinfo{year}{2023}\natexlab{}.
\newblock \showarticletitle{Understanding Design Collaboration Between Designers and Artificial Intelligence: A Systematic Literature Review}.
\newblock \bibinfo{journal}{\emph{Proceedings of the ACM on Human-Computer Interaction}} \bibinfo{volume}{7}, \bibinfo{number}{CSCW2} (\bibinfo{year}{2023}), \bibinfo{pages}{1--35}.
\newblock


\bibitem[Subramonyam et~al\mbox{.}(2021)]%
        {subramonyam2021towards}
\bibfield{author}{\bibinfo{person}{Hariharan Subramonyam}, \bibinfo{person}{Colleen Seifert}, {and} \bibinfo{person}{Eytan Adar}.} \bibinfo{year}{2021}\natexlab{}.
\newblock \showarticletitle{Towards a process model for co-creating AI experiences}. In \bibinfo{booktitle}{\emph{Designing Interactive Systems Conference 2021}}. \bibinfo{pages}{1529--1543}.
\newblock


\bibitem[Vaswani et~al\mbox{.}(2017)]%
        {vaswani2017attention}
\bibfield{author}{\bibinfo{person}{Ashish Vaswani}, \bibinfo{person}{Noam Shazeer}, \bibinfo{person}{Niki Parmar}, \bibinfo{person}{Jakob Uszkoreit}, \bibinfo{person}{Llion Jones}, \bibinfo{person}{Aidan~N Gomez}, \bibinfo{person}{{\L}ukasz Kaiser}, {and} \bibinfo{person}{Illia Polosukhin}.} \bibinfo{year}{2017}\natexlab{}.
\newblock \showarticletitle{Attention is all you need}.
\newblock \bibinfo{journal}{\emph{Advances in neural information processing systems}}  \bibinfo{volume}{30} (\bibinfo{year}{2017}).
\newblock


\bibitem[Wang et~al\mbox{.}(2023)]%
        {wang2023modelscope}
\bibfield{author}{\bibinfo{person}{Jiuniu Wang}, \bibinfo{person}{Hangjie Yuan}, \bibinfo{person}{Dayou Chen}, \bibinfo{person}{Yingya Zhang}, \bibinfo{person}{Xiang Wang}, {and} \bibinfo{person}{Shiwei Zhang}.} \bibinfo{year}{2023}\natexlab{}.
\newblock \showarticletitle{ModelScope Text-to-Video Technical Report}.
\newblock \bibinfo{journal}{\emph{arXiv preprint arXiv:2308.06571}} (\bibinfo{year}{2023}).
\newblock


\bibitem[Wei et~al\mbox{.}(2022)]%
        {wei2022emergent}
\bibfield{author}{\bibinfo{person}{Jason Wei}, \bibinfo{person}{Yi Tay}, \bibinfo{person}{Rishi Bommasani}, \bibinfo{person}{Colin Raffel}, \bibinfo{person}{Barret Zoph}, \bibinfo{person}{Sebastian Borgeaud}, \bibinfo{person}{Dani Yogatama}, \bibinfo{person}{Maarten Bosma}, \bibinfo{person}{Denny Zhou}, \bibinfo{person}{Donald Metzler}, {et~al\mbox{.}}} \bibinfo{year}{2022}\natexlab{}.
\newblock \showarticletitle{Emergent abilities of large language models}.
\newblock \bibinfo{journal}{\emph{arXiv preprint arXiv:2206.07682}} (\bibinfo{year}{2022}).
\newblock


\bibitem[Weintrop and Wilensky(2015)]%
        {weintrop2015block}
\bibfield{author}{\bibinfo{person}{David Weintrop} {and} \bibinfo{person}{Uri Wilensky}.} \bibinfo{year}{2015}\natexlab{}.
\newblock \showarticletitle{To block or not to block, that is the question: students' perceptions of blocks-based programming}. In \bibinfo{booktitle}{\emph{Proceedings of the 14th international conference on interaction design and children}}. \bibinfo{pages}{199--208}.
\newblock


\bibitem[Whitley and Blackwell(1997)]%
        {whitley1997visual}
\bibfield{author}{\bibinfo{person}{Kirsten~N Whitley} {and} \bibinfo{person}{Alan~F Blackwell}.} \bibinfo{year}{1997}\natexlab{}.
\newblock \showarticletitle{Visual programming: the outlook from academia and industry}. In \bibinfo{booktitle}{\emph{Papers presented at the seventh workshop on Empirical studies of programmers}}. \bibinfo{pages}{180--208}.
\newblock


\bibitem[Williams et~al\mbox{.}(2022)]%
        {williams2022ml}
\bibfield{author}{\bibinfo{person}{Randi Williams}, \bibinfo{person}{Micha{\l} Moskal}, {and} \bibinfo{person}{Peli De~Halleux}.} \bibinfo{year}{2022}\natexlab{}.
\newblock \showarticletitle{ML Blocks: A Block-Based, Graphical User Interface for Creating TinyML Models}. In \bibinfo{booktitle}{\emph{2022 IEEE Symposium on Visual Languages and Human-Centric Computing (VL/HCC)}}. IEEE, \bibinfo{pages}{1--5}.
\newblock


\bibitem[Wu et~al\mbox{.}(2023)]%
        {wu2023visual}
\bibfield{author}{\bibinfo{person}{Chenfei Wu}, \bibinfo{person}{Shengming Yin}, \bibinfo{person}{Weizhen Qi}, \bibinfo{person}{Xiaodong Wang}, \bibinfo{person}{Zecheng Tang}, {and} \bibinfo{person}{Nan Duan}.} \bibinfo{year}{2023}\natexlab{}.
\newblock \showarticletitle{Visual chatgpt: Talking, drawing and editing with visual foundation models}.
\newblock \bibinfo{journal}{\emph{arXiv preprint arXiv:2303.04671}} (\bibinfo{year}{2023}).
\newblock


\bibitem[Wu et~al\mbox{.}(2022)]%
        {wu2022promptchainer}
\bibfield{author}{\bibinfo{person}{Tongshuang Wu}, \bibinfo{person}{Ellen Jiang}, \bibinfo{person}{Aaron Donsbach}, \bibinfo{person}{Jeff Gray}, \bibinfo{person}{Alejandra Molina}, \bibinfo{person}{Michael Terry}, {and} \bibinfo{person}{Carrie~J Cai}.} \bibinfo{year}{2022}\natexlab{}.
\newblock \showarticletitle{Promptchainer: Chaining large language model prompts through visual programming}. In \bibinfo{booktitle}{\emph{CHI Conference on Human Factors in Computing Systems Extended Abstracts}}. \bibinfo{pages}{1--10}.
\newblock


\bibitem[Yang(2018)]%
        {yang2018machine}
\bibfield{author}{\bibinfo{person}{Qian Yang}.} \bibinfo{year}{2018}\natexlab{}.
\newblock \showarticletitle{Machine learning as a UX design material: how can we imagine beyond automation, recommenders, and reminders?}. In \bibinfo{booktitle}{\emph{AAAI Spring Symposia}}, Vol.~\bibinfo{volume}{1}. \bibinfo{pages}{2--6}.
\newblock


\bibitem[Yang et~al\mbox{.}(2018)]%
        {yang2018investigating}
\bibfield{author}{\bibinfo{person}{Qian Yang}, \bibinfo{person}{Alex Scuito}, \bibinfo{person}{John Zimmerman}, \bibinfo{person}{Jodi Forlizzi}, {and} \bibinfo{person}{Aaron Steinfeld}.} \bibinfo{year}{2018}\natexlab{}.
\newblock \showarticletitle{Investigating how experienced UX designers effectively work with machine learning}. In \bibinfo{booktitle}{\emph{Proceedings of the 2018 designing interactive systems conference}}. \bibinfo{pages}{585--596}.
\newblock


\bibitem[Yang et~al\mbox{.}(2020)]%
        {yang2020re}
\bibfield{author}{\bibinfo{person}{Qian Yang}, \bibinfo{person}{Aaron Steinfeld}, \bibinfo{person}{Carolyn Ros{\'e}}, {and} \bibinfo{person}{John Zimmerman}.} \bibinfo{year}{2020}\natexlab{}.
\newblock \showarticletitle{Re-examining whether, why, and how human-AI interaction is uniquely difficult to design}. In \bibinfo{booktitle}{\emph{Proceedings of the 2020 chi conference on human factors in computing systems}}. \bibinfo{pages}{1--13}.
\newblock


\bibitem[Yildirim et~al\mbox{.}(2022)]%
        {yildirim2022experienced}
\bibfield{author}{\bibinfo{person}{Nur Yildirim}, \bibinfo{person}{Alex Kass}, \bibinfo{person}{Teresa Tung}, \bibinfo{person}{Connor Upton}, \bibinfo{person}{Donnacha Costello}, \bibinfo{person}{Robert Giusti}, \bibinfo{person}{Sinem Lacin}, \bibinfo{person}{Sara Lovic}, \bibinfo{person}{James~M O'Neill}, \bibinfo{person}{Rudi~O'Reilly Meehan}, {et~al\mbox{.}}} \bibinfo{year}{2022}\natexlab{}.
\newblock \showarticletitle{How Experienced Designers of Enterprise Applications Engage AI as a Design Material}. In \bibinfo{booktitle}{\emph{Proceedings of the 2022 CHI Conference on Human Factors in Computing Systems}}. \bibinfo{pages}{1--13}.
\newblock


\bibitem[Yildirim et~al\mbox{.}(2023)]%
        {yildirim2023creating}
\bibfield{author}{\bibinfo{person}{Nur Yildirim}, \bibinfo{person}{Changhoon Oh}, \bibinfo{person}{Deniz Sayar}, \bibinfo{person}{Kayla Brand}, \bibinfo{person}{Supritha Challa}, \bibinfo{person}{Violet Turri}, \bibinfo{person}{Nina Crosby~Walton}, \bibinfo{person}{Anna~Elise Wong}, \bibinfo{person}{Jodi Forlizzi}, \bibinfo{person}{James McCann}, {and} \bibinfo{person}{John Zimmerman}.} \bibinfo{year}{2023}\natexlab{}.
\newblock \showarticletitle{Creating Design Resources to Scaffold the Ideation of AI Concepts}. In \bibinfo{booktitle}{\emph{Proceedings of the 2023 ACM Designing Interactive Systems Conference}} (Pittsburgh, PA, USA) \emph{(\bibinfo{series}{DIS '23})}. \bibinfo{publisher}{Association for Computing Machinery}, \bibinfo{address}{New York, NY, USA}, \bibinfo{pages}{2326–2346}.
\newblock
\showISBNx{9781450398930}
\urldef\tempurl%
\url{https://doi.org/10.1145/3563657.3596058}
\showDOI{\tempurl}


\bibitem[Zamfirescu-Pereira et~al\mbox{.}(2023)]%
        {zamfirescu2023johnny}
\bibfield{author}{\bibinfo{person}{JD Zamfirescu-Pereira}, \bibinfo{person}{Richmond~Y Wong}, \bibinfo{person}{Bjoern Hartmann}, {and} \bibinfo{person}{Qian Yang}.} \bibinfo{year}{2023}\natexlab{}.
\newblock \showarticletitle{Why Johnny can’t prompt: how non-AI experts try (and fail) to design LLM prompts}. In \bibinfo{booktitle}{\emph{Proceedings of the 2023 CHI Conference on Human Factors in Computing Systems}}. \bibinfo{pages}{1--21}.
\newblock


\bibitem[Zeng et~al\mbox{.}(2022)]%
        {zeng2022socratic}
\bibfield{author}{\bibinfo{person}{Andy Zeng}, \bibinfo{person}{Maria Attarian}, \bibinfo{person}{Brian Ichter}, \bibinfo{person}{Krzysztof Choromanski}, \bibinfo{person}{Adrian Wong}, \bibinfo{person}{Stefan Welker}, \bibinfo{person}{Federico Tombari}, \bibinfo{person}{Aveek Purohit}, \bibinfo{person}{Michael Ryoo}, \bibinfo{person}{Vikas Sindhwani}, {et~al\mbox{.}}} \bibinfo{year}{2022}\natexlab{}.
\newblock \showarticletitle{Socratic models: Composing zero-shot multimodal reasoning with language}.
\newblock \bibinfo{journal}{\emph{arXiv preprint arXiv:2204.00598}} (\bibinfo{year}{2022}).
\newblock


\bibitem[Zhang et~al\mbox{.}(2023)]%
        {zhang2023adding}
\bibfield{author}{\bibinfo{person}{Lvmin Zhang}, \bibinfo{person}{Anyi Rao}, {and} \bibinfo{person}{Maneesh Agrawala}.} \bibinfo{year}{2023}\natexlab{}.
\newblock \bibinfo{title}{Adding Conditional Control to Text-to-Image Diffusion Models}.
\newblock
\newblock


\bibitem[Zhou et~al\mbox{.}(2022)]%
        {zhou2022detecting}
\bibfield{author}{\bibinfo{person}{Xingyi Zhou}, \bibinfo{person}{Rohit Girdhar}, \bibinfo{person}{Armand Joulin}, \bibinfo{person}{Philipp Kr{\"a}henb{\"u}hl}, {and} \bibinfo{person}{Ishan Misra}.} \bibinfo{year}{2022}\natexlab{}.
\newblock \showarticletitle{Detecting twenty-thousand classes using image-level supervision}. In \bibinfo{booktitle}{\emph{European Conference on Computer Vision}}. Springer, \bibinfo{pages}{350--368}.
\newblock


\bibitem[Zimmerman et~al\mbox{.}(2007)]%
        {zimmerman2007research}
\bibfield{author}{\bibinfo{person}{John Zimmerman}, \bibinfo{person}{Jodi Forlizzi}, {and} \bibinfo{person}{Shelley Evenson}.} \bibinfo{year}{2007}\natexlab{}.
\newblock \showarticletitle{Research through design as a method for interaction design research in HCI}. In \bibinfo{booktitle}{\emph{Proceedings of the SIGCHI conference on Human factors in computing systems}}. \bibinfo{pages}{493--502}.
\newblock


\end{thebibliography}


\clearpage

\appendix
\onecolumn
\section{List of Implemented Models}
\label{appendix:models-list}

\begin{table}[H]
\caption{Full list of implemented models.}
\centering
\resizebox{\textwidth}{!}{%
\begin{tabular}{|l|l|l|l|}
\hline
\textbf{Puzzle piece name}                    & \textbf{Model name}    & \textbf{Input}              & \textbf{Output}       \\ \hline
Ask GPT                                       & GPT-4                  & Text                        & Text                  \\ \hline
Generate image                                & Stable Diffusion       & Text                        & Image                 \\ \hline
Generate video                                & AnimateDiff            & Text                        & Video                 \\ \hline
Generate 3D model                             & Shap-E                 & Text                        & 3D                    \\ \hline
Generate sound effects                        & AudioGen               & Text                        & Audio (Sound Effects) \\ \hline
Generate music                                & MusicGen               & Text                        & Audio (Music)         \\ \hline
Generate speech                               & Bark                   & Text                        & Audio (Speech)        \\ \hline
Describe image                                & BLIP                   & Image                       & Text                  \\ \hline
Tag image                                     & Segment Anything       & Image                       & Text                  \\ \hline
Extract text in image                         & PyTesseract            & Image                       & Text                  \\ \hline
Classify emotion from face                    & ResidualMaskingNetwork & Image                       & Text                  \\ \hline
Increase image resolution                     & Real-ESRGAN            & Image                       & Image                 \\ \hline
Restore distorted face                        & GFPGAN                 & Image                       & Image                 \\ \hline
Grayscale → Color                             & BigColor               & Image                       & Image                 \\ \hline
Remove image background                       & Rembg                  & Image                       & Image                 \\ \hline
Remove people                                 & EdgeConnect            & Image                       & Image                 \\ \hline
Get human pose                                & OpenPose               & Image                       & Image (Pose)          \\ \hline
Get segmentation map                          & Segment Anything       & Image                       & Image (Segmentation)  \\ \hline
Get depth map                                 & MiDaS                  & Image                       & Image (Depth)         \\ \hline
Get normal map                                & MiDaS                  & Image                       & Image (Normal)        \\ \hline
Get edge map                                  & HED                    & Image                       & Image (Edge)          \\ \hline
Generate 3D model from image                  & One-2-3-45             & Image                       & 3D                    \\ \hline
Classify video                                & X-CLIP                 & Video                       & Text                  \\ \hline
Remove video background                       & Robust Video Matting   & Video                       & Video                 \\ \hline
Increase video resolution                     & RealBasicVSR           & Video                       & Video                 \\ \hline
Increase video frame rate                     & RIFE                   & Video                       & Video                 \\ \hline
Classify music genre                          & MusicClassification    & Audio                       & Text                  \\ \hline
Transcribe speech                             & Whisper                & Audio                       & Text                  \\ \hline
Generate image from text and driving image    & Instruct-Pix2Pix       & Image + Text                & Image                 \\ \hline
Edit face with text                           & StyleCLIP              & Image + Text                & Image                 \\ \hline
Generate image from text and human pose       & ControlNet             & Image (Pose) + Text         & Image                 \\ \hline
Generate image from text and segmentation map & ControlNet             & Image (Segmentation) + Text & Image                 \\ \hline
Generate image from text and depth map        & ControlNet             & Image (Depth) + Text        & Image                 \\ \hline
Generate image from text and normal map       & ControlNet             & Image (Normal) + Text       & Image                 \\ \hline
Generate image from text and edge map         & ControlNet             & Image (Edge) + Text         & Image                 \\ \hline
Animate a face to talk                        & SadTalker              & Image + Audio               & Video                 \\ \hline
Clone a voice                                 & FreeVC                 & Audio + Audio               & Audio                 \\ \hline
Generate image from text and sketch           & ControlNet             & Sketch + Text               & Image                 \\ \hline
Generate artwork from text and sketch         & Scribble Stories       & Sketch + Text               & Image                 \\ \hline
\end{tabular}
}

\end{table}

\section{Participant Model Mosaics}
\label{appendix:user-chains}

\begin{table}[H]
\caption{Model mosaics created by participants during the free creation task.}
\resizebox{\columnwidth}{!}{%
\begin{tabular}{|l|l|l|l|}
\hline
\textbf{Participant} &
  \textbf{Free creation design} &
  \textbf{Names of puzzle pieces used} &
  \multicolumn{1}{l|}{\textbf{\begin{tabular}[c]{@{}l@{}}Number of\\ puzzle pieces\\ used\end{tabular}}} \\ \hline
P1 &
  Designing a formal dress &
  \begin{tabular}[c]{@{}l@{}}Ask GPT (custom, ideation, translation),\\ Tag image, Get human pose,\\ Generate image from text and pose,\\ Remove background, Generate image from text and image\end{tabular} &
  8 \\ \hline
P2 &
  Creating an audio-visual story &
  \begin{tabular}[c]{@{}l@{}}Ask GPT (ideation, translation),\\ Generate artwork from text and sketch,\\ Generate speech, Clone a voice, Generate sound effects\end{tabular} &
  6 \\ \hline
P3 &
  Creating a music video &
  \begin{tabular}[c]{@{}l@{}}Ask GPT (custom, ideation, translation),\\ Generate speech, Clone a voice, Classify music genre,\\ Generate music, Generate video, Increase video frame rate\end{tabular} &
  9 \\ \hline
P4 &
  Creating multiple movie endings &
  \begin{tabular}[c]{@{}l@{}}Ask GPT (custom, ideation, translation), Grayscale → Color,\\ Restore distorted face, Generate image, Generate video,\\ Generate music, Increase video resolution, Increase video frame rate\end{tabular} &
  10 \\ \hline
P5 &
  Designing a living room &
  \begin{tabular}[c]{@{}l@{}}Ask GPT (ideation, translation), Describe image,\\ Get segmentation map, Get depth map,\\ Generate image from text and segmentation map,\\ Generate image from text and depth map,\\ Remove background, Remove people\end{tabular} &
  9 \\ \hline
P6 &
  Designing a tote bag &
  \begin{tabular}[c]{@{}l@{}}Ask GPT (ideation, translation), Generate image,\\ Remove background, Generate image from text and image,\\ Get edge map, Generate image from text and edge map,\\ Generate artwork from text and sketch\end{tabular} &
  8 \\ \hline
P7 &
  Designing a logo &
  \begin{tabular}[c]{@{}l@{}}Ask GPT (custom, ideation, translation), Tag image,\\ Generate image, Generate image from text and sketch,\\ Generate artwork from text and sketch\end{tabular} &
  7 \\ \hline
P8 &
  Enhancing a story for the visually impaired &
  \begin{tabular}[c]{@{}l@{}}Ask GPT (ideation, translation), Extract text in image,\\ Describe image, Classify emotion from face, Generate speech,\\ Generate sound effects, Generate music\end{tabular} &
  8 \\ \hline
P9 &
  Designing a 3D model of a vase &
  \begin{tabular}[c]{@{}l@{}}Ask GPT (ideation, translation), Generate 3D model,\\ Generate image, Remove background, Generate 3D model from image\end{tabular} &
  6 \\ \hline
P10 &
  Designing a video game character &
  \begin{tabular}[c]{@{}l@{}}Ask GPT (ideation, translation), Generate image, Get human pose,\\ Generate image from text and human pose, Remove image background,\\ Generate speech, Animate a face to talk\end{tabular} &
  8 \\ \hline
\end{tabular}%
}

\end{table}

\newpage

\section{Semantic Search and Assembly Assistant Examples}
\label{appendix:subcomponent-examples}

\begin{table}[H]
\caption{\rev{Example search queries and returned relevant puzzle pieces using the Catalog Panel's semantic search.}}
\resizebox{\columnwidth}{!}{%
\begin{tabular}{|l|l|l|l|}
\hline
\textbf{Semantic search query} & \textbf{Returned relevant puzzle pieces}                                               \\ \hline
Improve content quality &
  \begin{tabular}[c]{@{}l@{}}Increase image resolution, Restore distorted face, Increase video resolution,\\ Increase video frame rate\end{tabular} \\ \hline
Describe the image             & Describe image, Tag image, Extract text in image, Classify emotion from face \\ \hline
Remove unwanted things         & Remove image background, Remove people, Remove video background              \\ \hline
Restore an old photo           & Grayscale → Color, Restore distorted face, Edit face with text               \\ \hline
Get the structure of an image  & Get depth map, Get edge map, Get normal map, Get segmentation map            \\ \hline
Generate an animation          & Generate video, Animate a face to talk                                       \\ \hline
Generate spoken words          & Generate speech, Clone a voice                                               \\ \hline
Generate content with only text &
  \begin{tabular}[c]{@{}l@{}}Ask GPT, Generate image, Generate video, Generate 3D model, Generate sound effects,\\ Generate music, Generate speech\end{tabular} \\ \hline
Generate content with an image reference &
  \begin{tabular}[c]{@{}l@{}}Generate image from text and driving image, Generate 3D model from image,\\ Edit face with text, Animate a face to talk\end{tabular} \\ \hline
\end{tabular}%
}

\end{table}

\begin{table}[H]
\caption{\rev{Example task queries and recommended combination of puzzle pieces using the Assembly Assistant.}}
\resizebox{\columnwidth}{!}{%
\begin{tabular}{|l|l|l|l|}
\hline
\textbf{Assembly Assistant task query}        & \textbf{Recommended combination of puzzle pieces}                                  \\ \hline
Design clothes for a model               & Get human pose, Ask GPT, Generate image from text and human pose \\ \hline
Create an animated cartoon character & Ask GPT, Generate image, Generate speech, Animate a face to talk          \\ \hline
Make a recipe for chocolate chip cookies & Ask GPT, Generate speech                                         \\ \hline
Add sound for a illustration book        & Describe image, Ask GPT, Generate sound effects                  \\ \hline
Create a podcast episode on finance      & Ask GPT, Generate speech, Clone a voice                          \\ \hline
Remodel my home interior             & Tag image, Ask GPT, Get depth map, Generate image from text and depth map \\ \hline
Create a short sci-fi film               & Ask GPT, Generate video, Generate speech, Generate sound effects \\ \hline
Plan a trip to San Francisco             & Ask GPT, Generate speech                                         \\ \hline
Turn my sketch into an artwork           & Ask GPT, Generate artwork from text and sketch                   \\ \hline
Give me gardening advice for my plants   & Tag image, Ask GPT                                               \\ \hline
\end{tabular}%
}

\end{table}

\end{document}